\documentclass[review]{elsarticle}

\usepackage{lineno,hyperref}

\usepackage{enumitem}
\usepackage{url}
\usepackage{glossaries}
\usepackage{graphicx}
\usepackage{subfig}
\usepackage{color}
\usepackage{todonotes}
\usepackage{ulem}
\usepackage{amssymb}

\modulolinenumbers[1]

\usepackage[utf8x]{inputenc}

\usepackage[T1]{fontenc}

\journal{Astroparticle Physics}

\begin{document}

\newcommand{\mcprod}[1]{\texttt{prod#1}}
\newacronym{gf}{GF}{Geomagnetic Field}
\newacronym{mc}{MC}{Monte Carlo}
\newacronym{eas}{EAS}{Extensive Air Showers}
\newacronym{cta}{CTA}{Cherenkov Telescope Array}
\newacronym{nsb}{NSB}{night-sky background}

\binoppenalty=10000
\relpenalty=10000

\newcommand{\tmtexttt}[1]{{\ttfamily{#1}}}
\newcommand{\apj}{Astrophysical Journal}
\newcommand{\aap}{A\&A}

\definecolor{navyblue}{rgb}{0.0, 0.0, 0.5}
\newcommand{\thadd}[1]{\textcolor{navyblue}{[TH: #1]}}

\begin{frontmatter}

\title{Monte Carlo studies for the optimisation of the Cherenkov Telescope Array layout
}


\author[1]{A.~Acharyya}
\author[2]{I.~Agudo}
\author[57]{E.O.~Angüner}
\author[3]{R.~Alfaro}
\author[4]{J.~Alfaro}
\author[5]{C.~Alispach}
\author[6]{R.~Aloisio}
\author[7]{R.~Alves Batista}
\author[8]{J.-P.Amans}
\author[9]{L.~Amati}
% ...

\address[1]{Dept. of Physics and Centre for Advanced Instrumentation, Durham University, South Road, Durham DH1 3LE, United Kingdom}
\address[2]{Instituto de Astrofísica de Andalucía-CSIC, Glorieta de la Astronomía s/n, E-18008, Granada, Spain}
\address[3]{Universidad Nacional Autónoma de México, Delegación Coyoacán, 04510 Ciudad de México, Mexico}
\address[4]{Pontificia Universidad Católica de Chile, Avda. Libertador Bernardo O' Higgins No 340, borough and city of Santiago, Chile}
\address[5]{University of Geneva - Département de physique nucléaire et corpusculaire, 24 rue du Général-Dufour, 1211 Genève 4, Switzerland}
\address[6]{INFN Dipartimento di Scienze Fisiche e Chimiche - Università degli Studi dell'Aquila and Gran Sasso Science Institute, Via Vetoio 1, Viale Crispi 7, 67100 L'Aquila, Italy}
\address[7]{Instituto de Astronomia, Geofísica, e Ciências Atmosféricas - Universidade de São Paulo, Cidade Universitária, R. do Matão, 1226, CEP 05508-090, São Paulo, SP, Brazil}
\address[8]{LUTH and GEPI, Observatoire de Paris, CNRS, PSL Research University, 5 place Jules Janssen, 92190, Meudon, France}
\address[9]{INAF - Osservatorio di astrofisica e scienza dello spazio di Bologna, Via Piero Gobetti 101, 40129  Bologna, Italy}
% ...
\address[57]{Aix Marseille Univ, CNRS/IN2P3, CPPM, Marseille, France, 163 Avenue de Luminy, 13288 Marseille cedex 09, France}
% ...

\cortext[cor1]{Corresponding author}

\begin{abstract}

The Cherenkov Telescope Array (CTA) is the major next-generation observatory for ground-based very-high-energy gamma-ray astronomy. It will improve the sensitivity of current ground-based instruments by a factor of five to twenty, depending on the energy, greatly improving both their angular and energy resolutions over four decades in energy (from 20 GeV to 300 TeV). 
This achievement will be possible by using tens of imaging Cherenkov telescopes of three successive sizes. They will be arranged into two arrays, one per hemisphere, located on the La Palma island (Spain) and in Paranal (Chile).
We present here the optimised and final telescope arrays for both CTA sites, as well as their foreseen performance, resulting from the analysis of three different large-scale Monte Carlo productions.
\end{abstract}


\end{frontmatter}
\newpage

\glsreset{cta}

\section{Introduction}

Cosmic rays and very-high-energy (VHE, few tens of GeV and above) gamma rays reaching Earth's atmosphere produce cascades of subatomic particles called air showers. Ultrarelativistic charged particles generated within these showers produce photons through the Cherenkov effect. Most of this light is emitted at altitudes ranging between 5 to 15 km, and it propagates down to ground level as a quasi-planar, thin disk of Cherenkov photons orthogonal to the shower axis.

Imaging atmospheric Cherenkov telescopes (IACTs) are designed to capture images of these very brief optical flashes, generally lasting just a few ns. By placing arrays of IACTs within the projected light pool of these showers and analysing the simultaneous images taken by these telescopes, it is possible to identify the nature of the primary particle and reconstruct its original energy and incoming direction.

Building on the experience gained through the operation of the current IACTs (H.E.S.S.\footnote{\url{https://www.mpi-hd.mpg.de/hfm/HESS/HESS.shtml}.}, MAGIC\footnote{\url{https://magic.mpp.mpg.de/}.}, and VERITAS\footnote{\url{https://veritas.sao.arizona.edu/}.}), the next generation of ground-based very-high-energy gamma-ray telescope is currently under construction. 
The \gls{cta}\footnote{\url{http://www.cta-observatory.org/}.} \cite{CTA_concept, CTAICRC2015} will detect gamma rays in the energy range from 20 GeV to 300 TeV with unprecedented angular and energy resolutions for ground-based facilities, outperforming the sensitivity of present-day instruments by more than an order of magnitude in the multi-TeV range \cite{CTA_MC_2017}. This improvement will be possible by using larger arrays of telescopes. As a cost-effective solution to improve performance over four decades of energy, telescopes will be built in three different sizes: Large-Sized Telescopes (LSTs) \cite{LSTgamma},  Medium-Sized Telescopes (MSTs) \cite{MSTgamma, SCTgamma} and Small-Sized Telescopes (SSTs) \cite{Montaruli:2015}. To provide full-sky coverage, IACT arrays will be installed in two sites, one in each hemisphere: at Paranal (Chile) and at La Palma (Canary Islands, Spain).

Each telescope class will primarily cover a specific energy range:
LSTs, with a $\sim370$ m$^2$ reflecting dish and a camera with a field of view (FoV) of $\sim4.3^\circ$, will allow the reconstruction of the faint low-energy showers (below 100 GeV), not detectable by smaller telescopes. In this energy range the rejection of the cosmic-ray background is limited by the modest number of particles created in the air showers. Due to the relatively high flux of low-energy gamma rays and the large associated construction costs, few LSTs will be built at each site. They have been designed for high-speed slewing allowing short repositioning times to catch fast transient phenomena on time scales of minutes to days, such as gamma-ray bursts \cite{ScienceCTA}.

MSTs, with a larger FoV of $\sim7.6^\circ$, will populate the inner part of the array, increasing the number of telescopes simultaneously observing each shower, enhancing the angular and energy resolutions within the \gls{cta} core energy range (between 100 GeV and 10 TeV). Two different MST designs have been proposed: the Davies-Cotton MST (DC-MST) and the Schwarzschild-Couder MST (SC-MST) \cite{MSTgamma, SCTgamma}. The DC-MST is a 12m-diameter single-mirror IACT built with modified Davies-Cotton optics and a mirror area of $\sim88$~m$^2$. Two different cameras have been prototyped for this telescope: NectarCam and FlashCam \cite{nectar, flashcam}. The SC-MST features a two-mirror optical design with a 9.7~m diameter primary mirror and an area of $\sim41$m$^2$. The dual-mirror setup corrects spherical and comatic aberrations, allowing a finer shower image pixelisation, enhancing angular resolution and off-axis performance.

Above a few TeV, Cherenkov light from electromagnetic showers becomes significantly brighter, not requiring such large reflecting surfaces for their detection. At the same time, the gamma-ray flux decreases with energy, so in order to detect a sufficient number of these high-energy events, a large ground surface needs to be covered. SSTs, with a mirror area of $\sim$ 8 m$^2$ and a FoV of $>8^\circ$, have been designed with this purpose. A large number of SSTs will populate the outer part of the array covering a total surface area of up to 4.5 km$^2$. Three variants of SSTs have been proposed: two designs of SC-SSTs, the ASTRI and the GCT, both with primary mirror diameters of 4~m, and a DC-SST, the SST-1M, with a single 4 m diameter mirror \cite{Montaruli:2015}.

The northern and southern observatories will make the full VHE gamma-ray sky accessible to CTA. As a cost-effective solution to maximise scientific output, each site will have different telescope layouts. The CTA southern site will be larger to take advantage of its privileged location for observation of the Galactic Center and most of the inner half of the Galactic Plane, regions with a high density of sources with spectra extending beyond 10 TeV. Its baseline design foresees 4 LSTs, 25 MSTs and 70 SSTs. The northern site  
will be more focused on the study of extragalactic objects 
and will be composed of 4 LSTs and 15 MSTs. No SSTs are planned to be placed in the northern hemisphere.

Detailed \gls{mc} simulations are required to estimate the performance of an IACT array \cite{Konrad:2008, APP_CTA_MC, MC_ICRC:2013}, which is evaluated by quantities like the minimum detectable flux, sensitive FoV or its angular and energy resolutions. All these estimators are strongly dependent on a set of parameters related to both the telescope design and the array layout (i.e. the arrangement of telescope positions on ground). Other scenarios (e.g. standalone operations of sub-arrays composed of only LSTs, MSTs or SSTs, or short downtime periods of some telescope) need to be also taken into consideration during the layout optimisation phase to ensure that the CTA performance is not critically affected. The objective of this work is to optimise the telescope layout of a given number of telescopes, maximising performance, while complying with all CTA requirements. These requirements were derived as a cost-effective solution to obtain excellent performance over a wide range of very different physics cases \cite{ScienceCTA}, to ensure the scientific impact of the future observatory.

\subsection{Array layout considerations}

Optimal array layouts are mainly characterised by the configuration of each telescope type and by the number and arrangement of these telescopes.
Each telescope type configuration is mainly described by its light collection power, dominated by mirror area, photo sensor efficiency, and camera FoV and pixelation, with optics chosen so that the optical point spread function matches the pixel size. A generic telescope cost model was used with mirror area, FoV and pixel size as primary parameters, so that all proposed array layouts that were compared during these optimisation studies could be considered of approximately equal cost.

As a first step, semi-analytical performance estimations were carried out using parameterisations for the responses 
of each telescope type. These studies allowed us to perform quick estimates of gamma-ray and cosmic-ray detection rates for a wide variety of telescope configurations and arrangements. Simulations of regular square grids of telescopes were performed to quantify the impact of parameters such as mirror area, FoV, pixel size or telescope spacing.

To validate and fine tune the optimal telescope configurations calculated with these simplified approaches, a series of large-scale MC simulations were performed sequentially, described in more detail in Section \ref{sec:mcprod}.

Telescopes are arranged in concentric arrays of different telescope sizes, ordered in light collection power, from a compact low-energy array at the centre to an extended high-energy array, providing an effective area that increases with energy. The light pool size of air showers increases with energy, from a radius of about 120 m for $\sim$ 30 GeV showers to more than 1000 m for multi-TeV showers. In the sub-TeV to TeV domain, telescope spacing of about 100 m to 150 m optimises sensitivity, providing an equilibrium between having more images per air shower and a reasonable collection area. For TeV energies and above, larger distances are preferred to improve the collection area, given that, at these energies, the cosmic-ray background can be rejected almost completely and the achievable sensitivity is photon-rate limited.

The baseline design number of telescopes (4 LSTs, 25 MSTs and 70 SSTs for CTA-South and 4 LSTs and 15 MSTs for CTA-North) was fixed after a combined effort involving the production of large-scale MC simulations, evaluation of the performance of very different array layouts \citep{APP_CTA_MC}, and study of the effect of this diverse set of layouts over a large variety of key scientific cases \cite{DM_2013, CRs_2013, AGNs_2013, surveys_2013, CTA_GRBs, pulsars_2013}.

This study presents the final baseline arrays for both the \gls{cta} northern and southern sites. \gls{mc} large-scale productions, described in section \ref{sec:mcprod}, were used to estimate the performance of a very large variety of layouts.
The main considerations taken into account in the performance evaluation are outlined in section \ref{sec:conlayout}, while the final baseline arrays and their performances are presented in sections \ref{sec:souths} and \ref{sec:norths} for the southern and northern site, respectively.

\section{CTA Monte Carlo production and analysis}
\label{sec:mcprod}

Given the unprecedented scale of the \gls{cta} project, a constant effort has been devoted over the past five years to define and optimise the telescope layouts. Three large-scale \gls{mc} productions were conducted and analysed with this purpose \cite{MC_ICRC:2013, Hassan-2015, ICRC2017:Layout}. In addition to the layout optimisation, these productions have been used to: 
\begin{itemize}
    \item estimate the expected \gls{cta} performance \cite{CTA_MC_2017, APP_CTA_MC},
    \item guide the design of the different telescope types and compare their capabilities \cite{Wood:2014, Prod2_SCMST, GCT_MC_2015},
    \item provide input to the site selection process by evaluating the effect of the characteristics of each site on the array performance. Among the considered site attributes there were altitude, geomagnetic field, night-sky background level and aerosol optical depth \cite{Site_Paper, MC_ICRC_site:2015, 2013APh....45....1S}.
\end{itemize}

As described in \cite{APP_CTA_MC,Site_Paper, Hassan-2015}, each large-scale \gls{mc} production requires the definition of a large telescope layout, called the master layout. Each master layout comprises hundreds of telescopes distributed over an area of about 6~km$^2$ and are designed to contain numerous possible \gls{cta} layouts of equivalent cost. To identify the optimal arrangement, these plausible layouts are extracted, analysed and their performances are compared with respect to each other. For each MC production, telescope models were sequentially improved, becoming more realistic in each iteration thanks to the increasing input coming from the prototype telescopes. Air showers initiated by gamma rays, cosmic-ray nuclei and electrons are simulated using the \tmtexttt{CORSIKA} package \cite{corsika}. 
The telescope response is simulated using \tmtexttt{sim\_telarray} \cite{Konrad:2008}, used by the HEGRA and H.E.S.S. experiments.

The simulated products generated by these large-scale productions resemble the data that will be supplied by the future CTA hardware and software. The performance of each telescope layout is estimated by analysing these data products using reconstruction methods \cite{evnDisplay, MARS}, developed for the current generation of IACTs, and adapted for analysis of the \gls{cta} arrays, briefly described in section \ref{sec:anchain}.

The first large-scale production (\mcprod{1}) covered a wide range of different layouts \cite{APP_CTA_MC}, from very compact ones, focused on low energies, to very extended ones, focused on multi-TeV energies. The evaluation of these layouts, studying their impact on a range of science cases \cite{DM_2013, CRs_2013, AGNs_2013, surveys_2013, CTA_GRBs, pulsars_2013}, resulted in a clear preference for intermediate layouts with a balanced performance over a wide energy range. 

The second large-scale production (\mcprod{2}) refined the layout optimisation studies \cite{Hassan-2015} while putting an additional emphasis on assessing the effect of site-related parameters over performance at the proposed sites to host the \gls{cta} Observatory \cite{MC_ICRC_site:2015}. Results from this production concluded that all proposed sites were excellent candidates to host CTA, but that sites at moderate altitudes ($\sim2000$ m) give the best overall performances \cite{Site_Paper}. Given the wide scope of this production, the layout optimisation performed \cite{Hassan-2015} is estimated to be $\sim10$\% away from the optimum 
performance, mainly due to the limited number of simulated telescope positions for a given site.

The third large-scale production (\mcprod{3}) was carried out for the primary \gls{cta} site candidates, Paranal (Chile) and La Palma (Spain). Telescope design configurations were updated and a significantly larger and more realistic set of available telescope positions were included (see Fig. \ref{fig:Layout_ALL_Prod3}). The aim of this production was to refine the optimisation, defining the final telescope layout for both \gls{cta} arrays by reducing the optimisation uncertainty to the few percent level, while preserving the goal of a balanced intermediate layout fulfilling all \gls{cta} performance requirements. 
To validate the baseline arrays inferred from this work (see section \ref{sec:souths}), this production was extended using identical telescope models. Telescope locations were further refined by considering a total of 210 positions for Paranal. All results presented in this paper, unless otherwise stated, refer to this third large-scale production.

The optimisation of the \gls{cta} arrays required a significant computational effort: the third large-scale production for the Paranal site alone required \hbox{$\approx$ 120} million HEP-SPEC06 CPU hours\footnote{The HEP-wide benchmark for measuring CPU performance. See specifications in \hbox{\url{http://w3.hepix.org/benchmarks}.}} and \hbox{$\approx$ 1.4} PB of disk storage. Most of these simulations were carried out on the CTA computing grid, using the European Grid Infrastructure and utilising the DIRAC framework as interware \cite{dirac-general, Arrabito-2015}, as well as on the computer clusters of the  Max-Planck-Institut f\"ur Kernphysik. The subsequent analysis was carried out using the DIRAC framework, as well as the computing clusters at the Deutsches Elektronen-Synchrotron and at the Port d'Informaci\'{o} Cient\'{i}fica.

\subsection{Simulated telescope layouts}\label{SimTelLay}

\begin{figure}[hpt]
\begin{center}
\includegraphics[height=0.4\textheight]{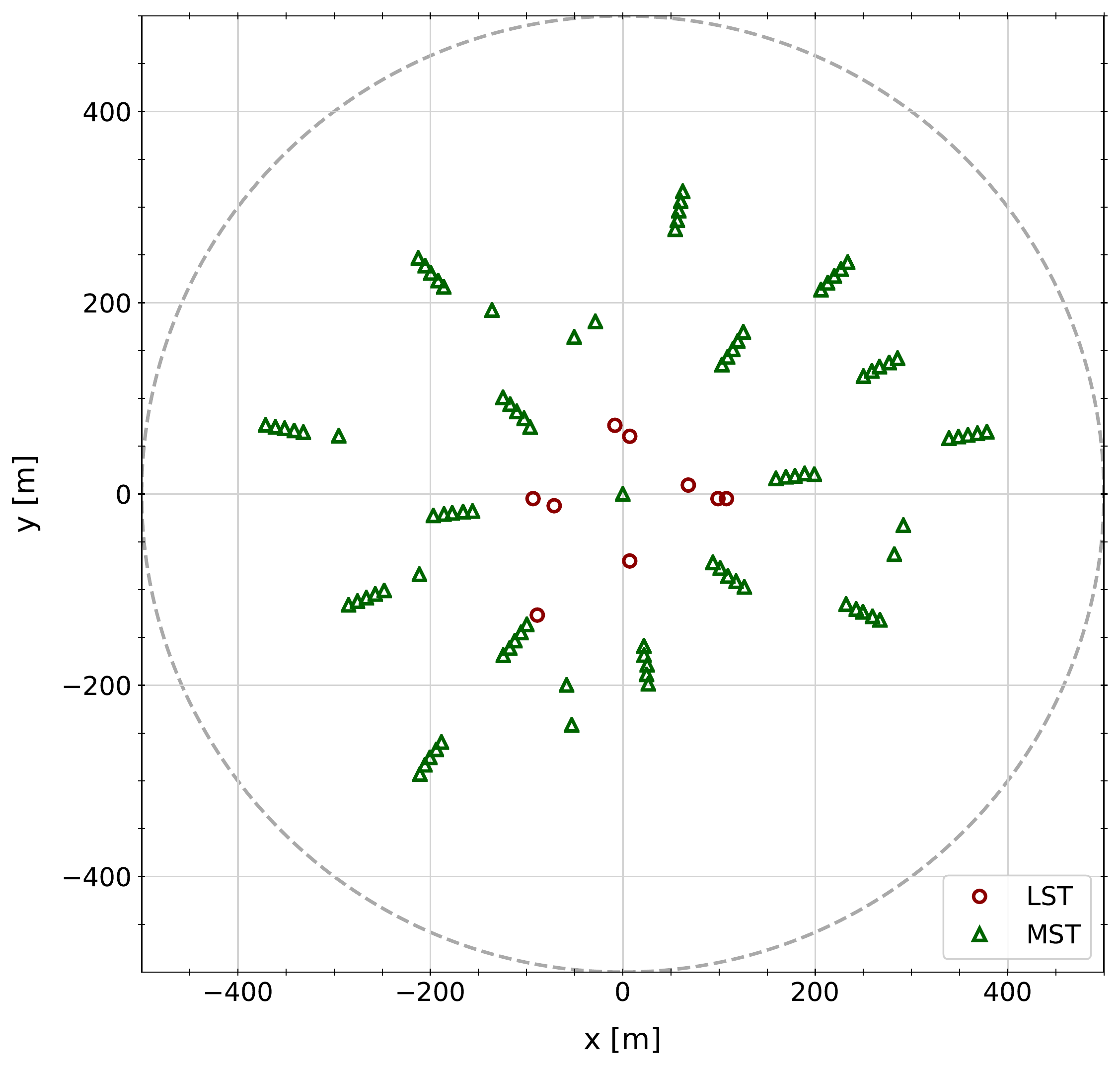}
\includegraphics[height=0.4\textheight]{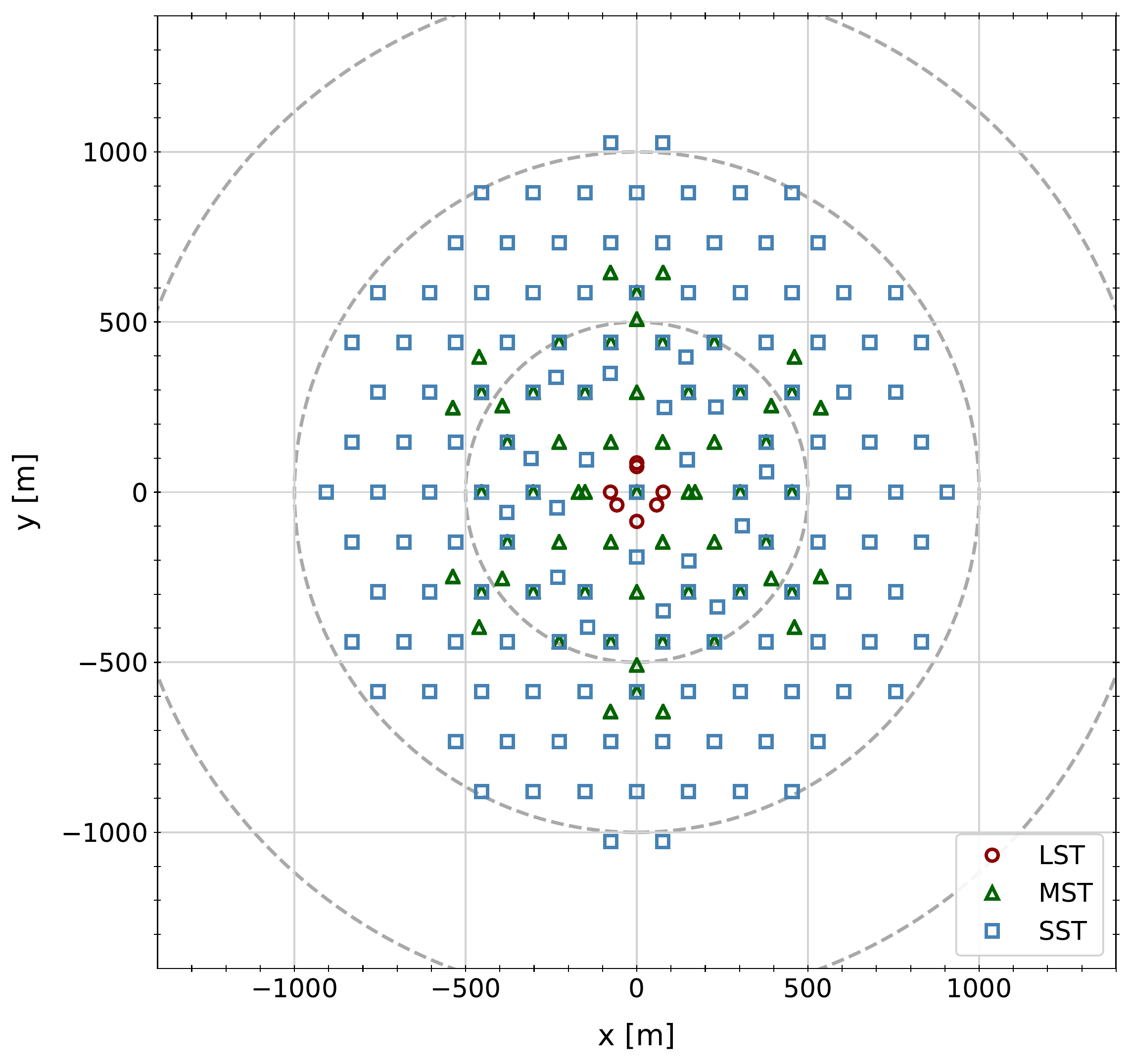}
\caption{Simulated telescope positions within the third large-scale MC production (see section \ref{sec:mcprod} for details). \textit{Top}: La Palma telescope positions including all radially-scaled MST layouts. The available positions are restricted by the site topography, buildings and roads. \textit{Bottom}: Paranal telescope positions before applying any radially-symmetric transformation (scaling number 1). LST positions are indicated by red circles, MSTs by green triangles, and SSTs by blue squares.}
\label{fig:Layout_ALL_Prod3}
\end{center}
\end{figure}

\begin{figure}[p]
\begin{center}
\includegraphics[width=0.44\textwidth]{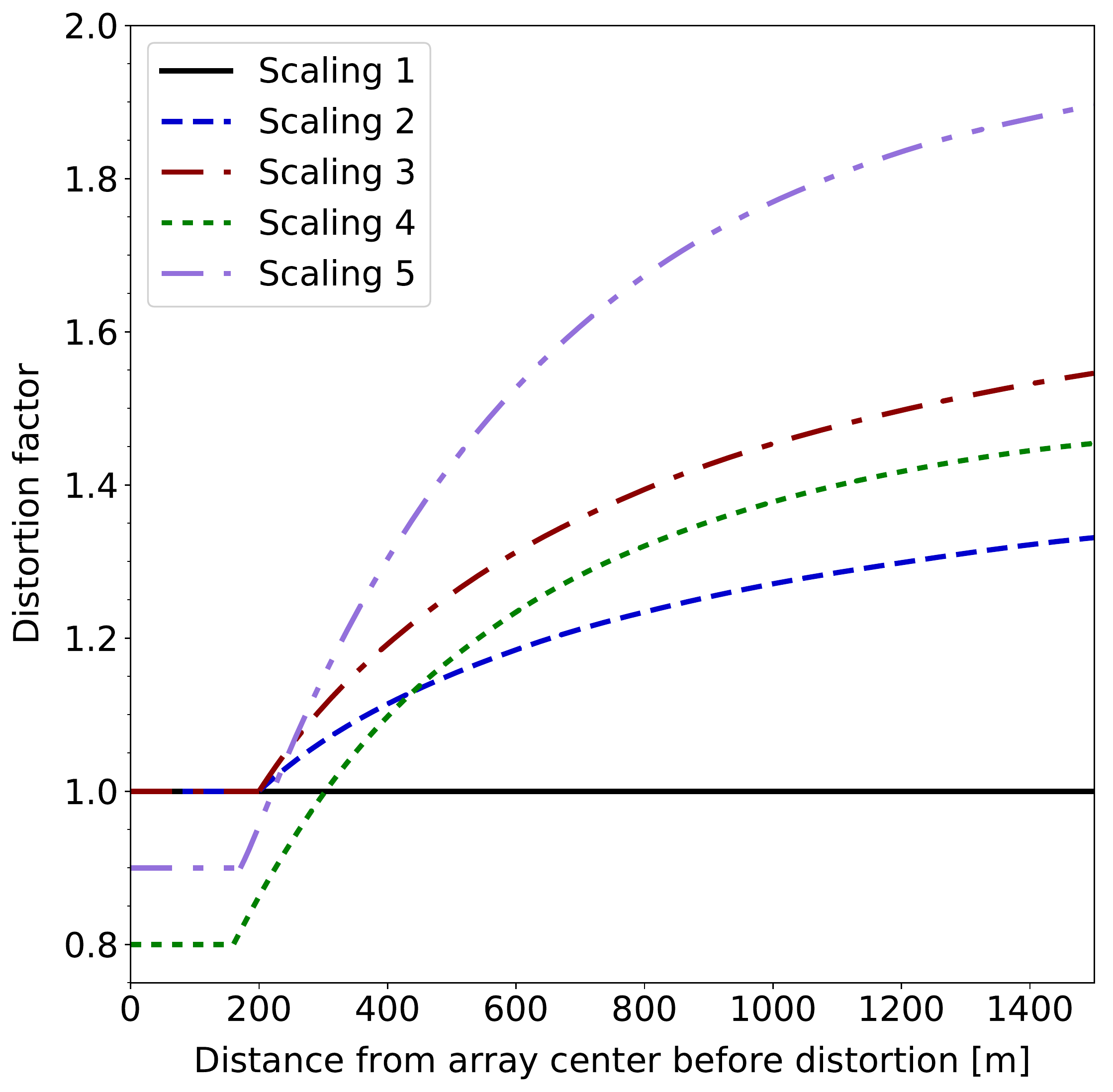}
\includegraphics[width=0.45\textwidth]{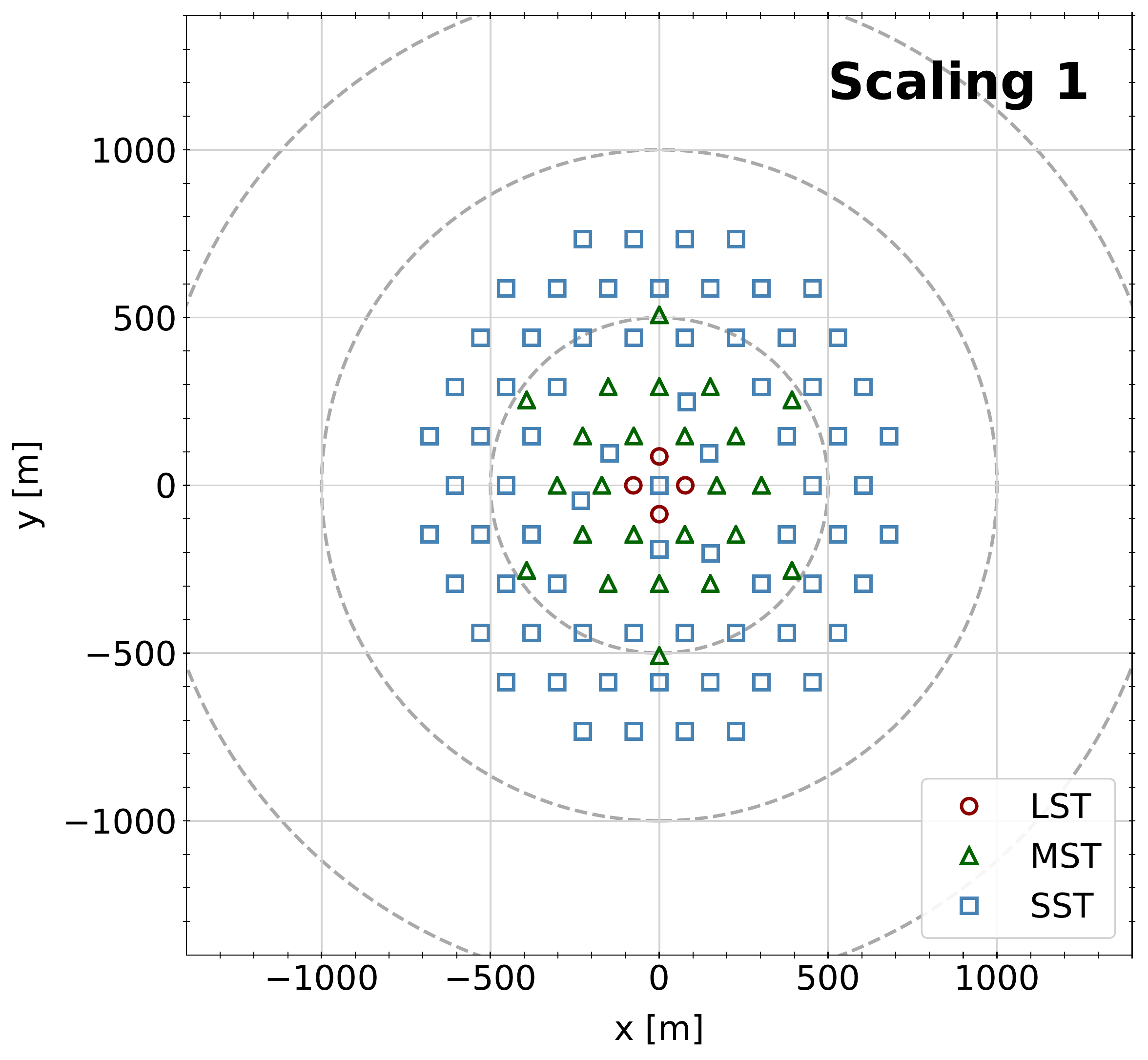}
\includegraphics[width=0.45\textwidth]{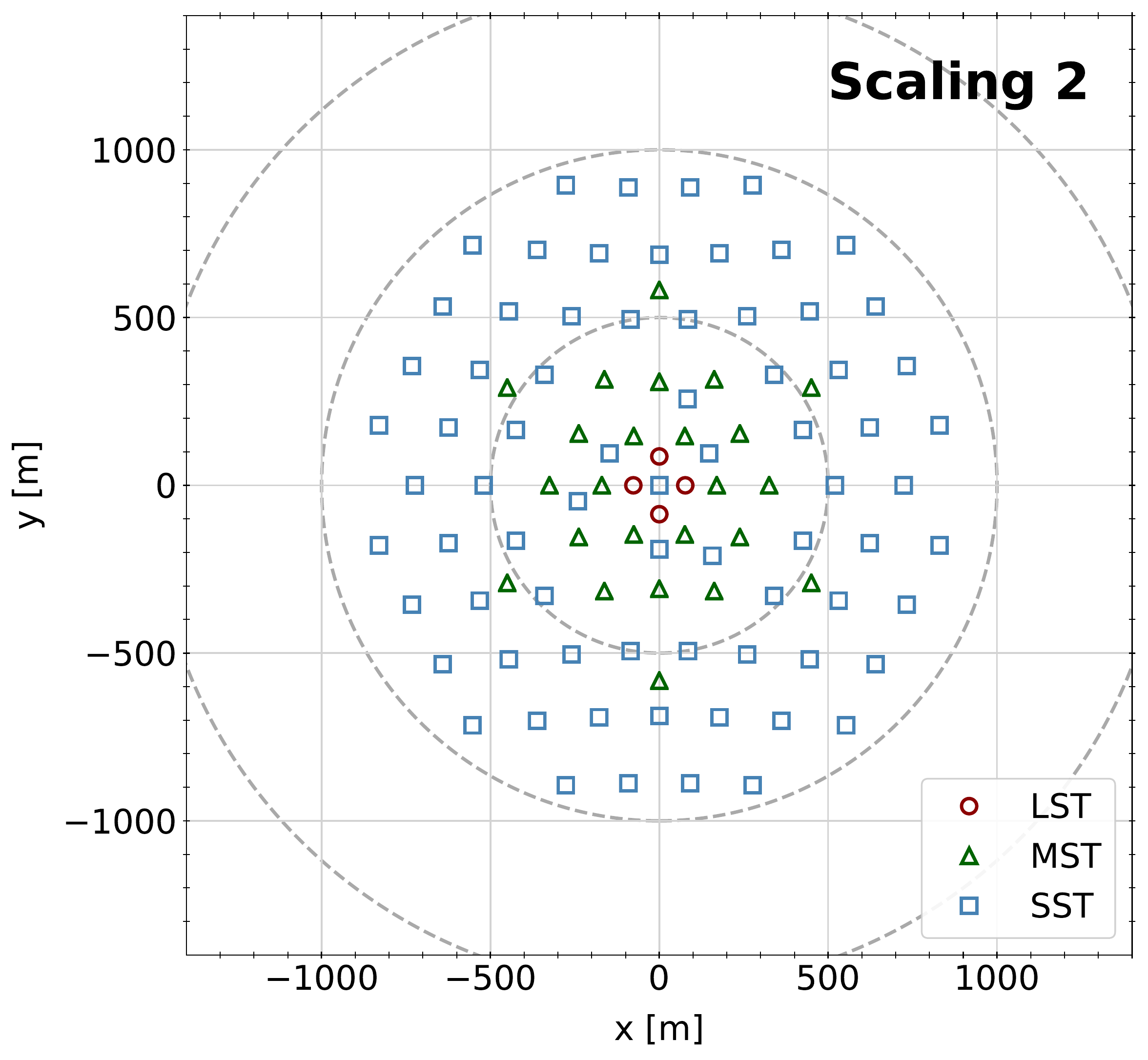}
\includegraphics[width=0.45\textwidth]{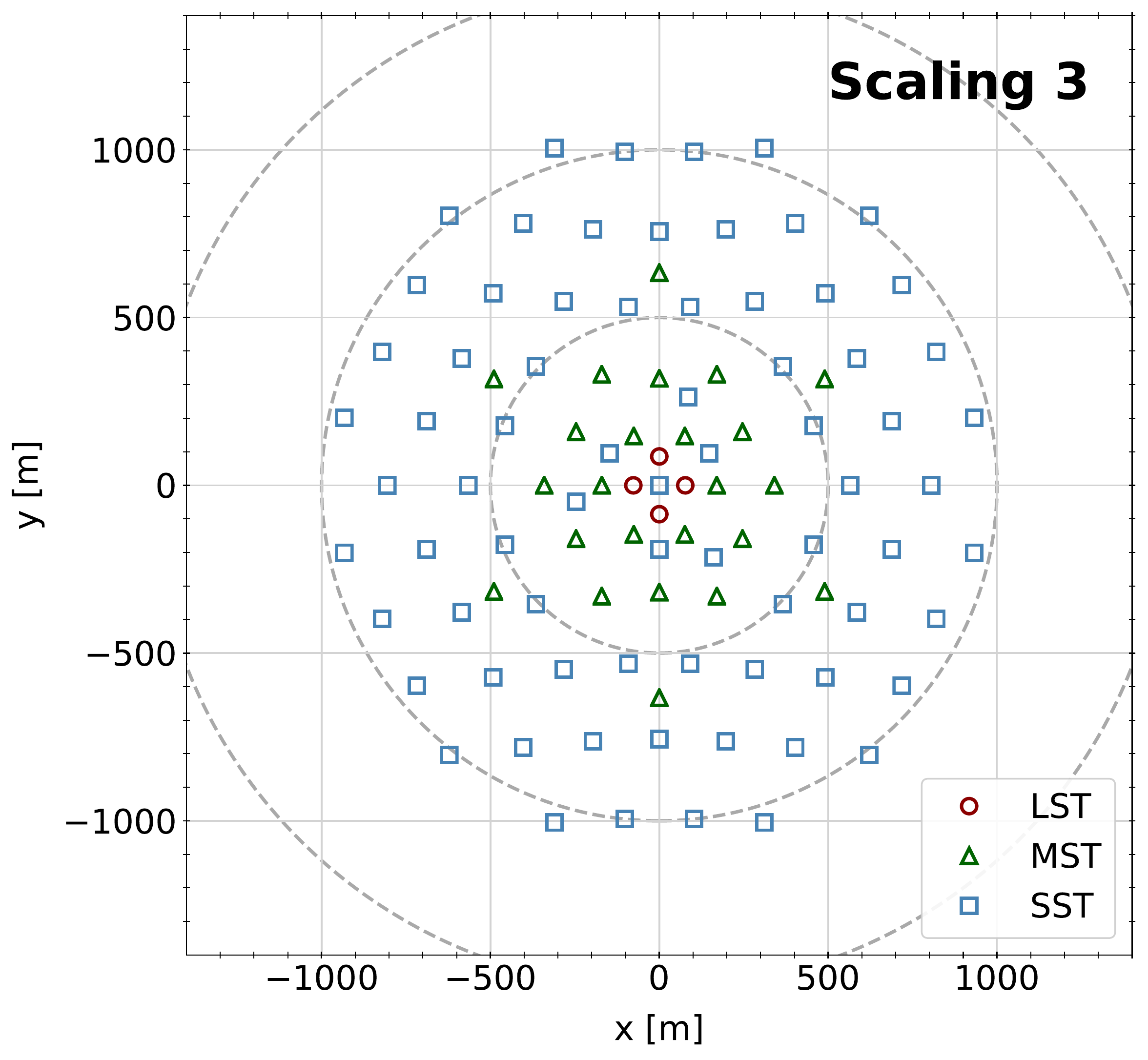}
\includegraphics[width=0.45\textwidth]{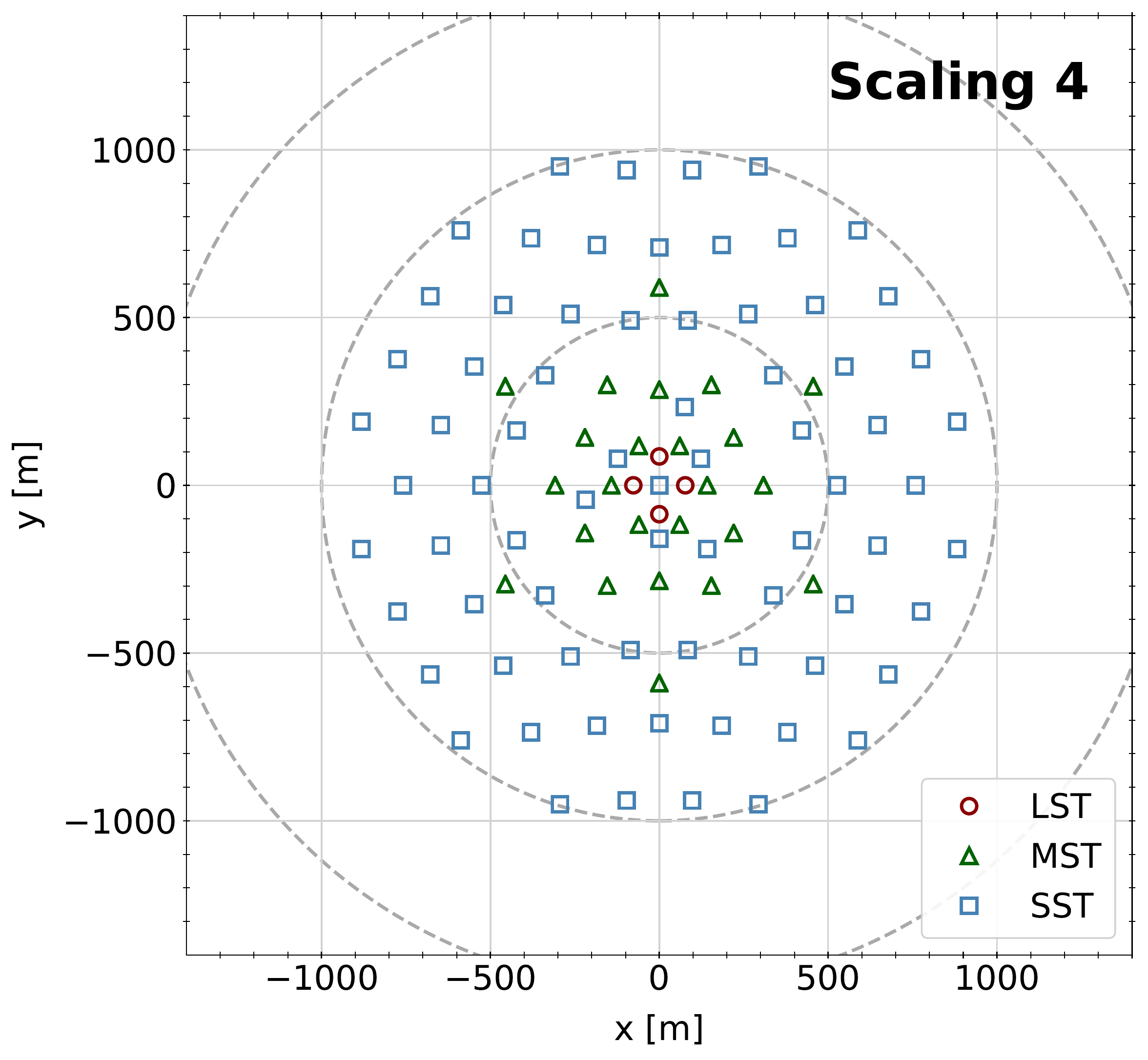}
\includegraphics[width=0.45\textwidth]{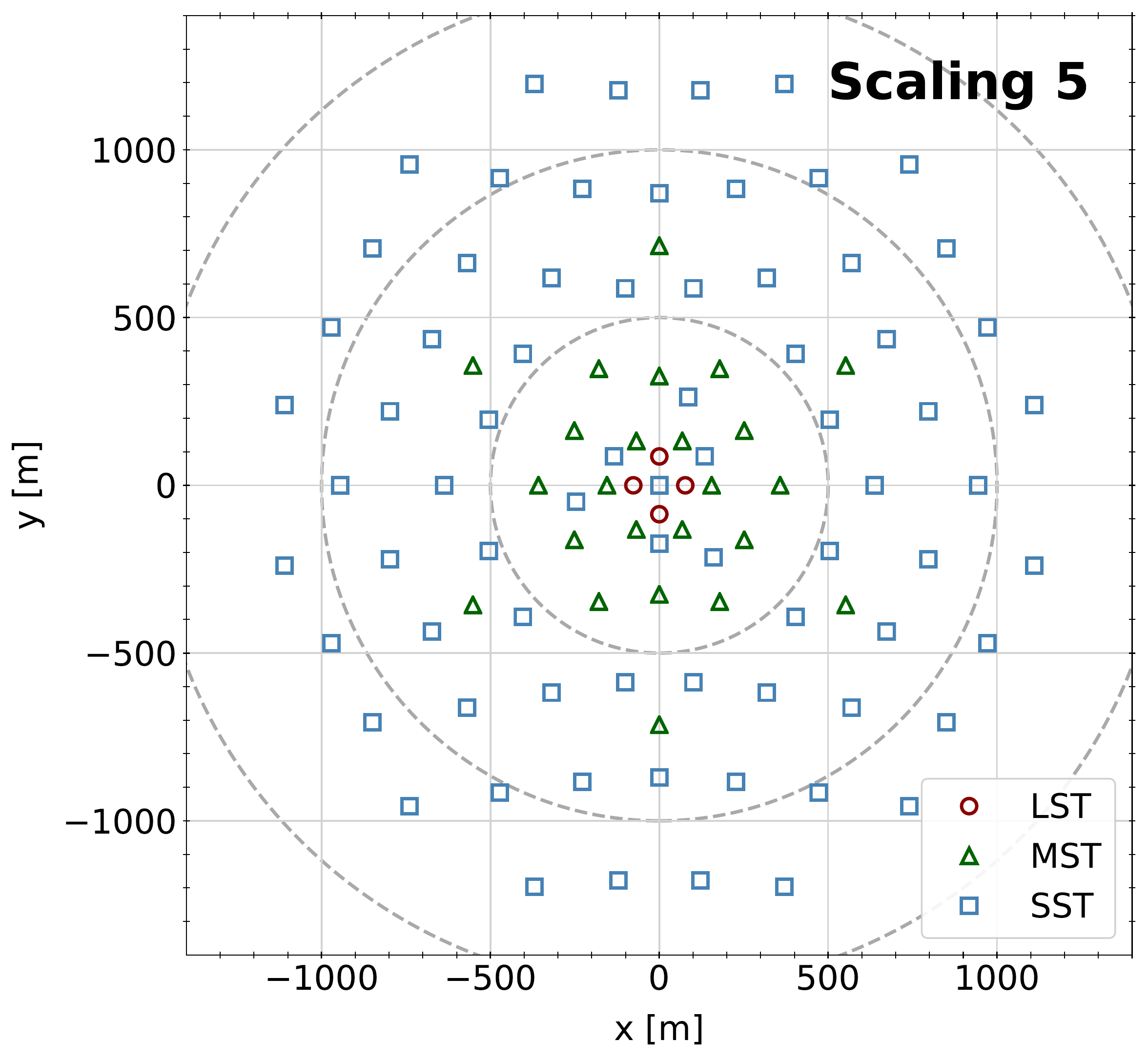}
\caption{\textit{Top-left}: Radially-symmetric distortion factors for the five different scalings applied to the CTA-South layouts, as a function of the radial distance to the centre of the array before the applied transformation. \textit{Top-right to bottom-right}: an example of the five resulting scaled layouts for one of the Paranal site candidates (``S1’’). LST positions are indicated by red circles, MSTs by green triangles, and SSTs by blue squares. Taken from \cite{ICRC2017:Layout}.}
\label{fig:distfactor}
\end{center}
\end{figure}

Layouts with a more compact and denser distribution of telescopes improve the direction and energy reconstruction of showers (the limiting factor for the low/mid-energy range of CTA, between 20 GeV and 5 TeV), while larger and sparser layouts improve the collection area and event statistics (the limiting factor for the highest energies), see also discussion in \cite{APP_CTA_MC}. To find the most efficient inter-telescope distance for CTA, each layout candidate is modified by applying several radially-symmetric scaling factors (see Fig. \ref{fig:distfactor}). On top of that, in order to maintain the radial symmetry of the array in the shower projection for typical observation directions near source culmination, the southern array layouts were stretched by a factor of 1.06 in the north-south direction and compressed by a factor 1/1.06 in the east-west direction. The assumption of an average culmination zenith angle of $z \sim 27^{\circ} \approx \arccos{(1/1.06^2)}$, is based on long-term observation statistics from H.E.S.S., MAGIC, and VERITAS.

The simulated telescope positions are shown in Figure \ref{fig:Layout_ALL_Prod3}. In the case of La Palma, for which a combination of all scaled layouts is shown, these positions were constrained by site topography, as well as by existing buildings and roads. For Paranal, the layout was based on a hexagonal grid\footnote{As discussed in \cite{Colin_2009}, a square grid is preferred to enhance two telescope events while a hexagonal layout favours the simultaneous detection of showers by three or more telescopes, the latter being more suitable for CTA.} with some additional positions. Five sets of radially-symmetric transformations were applied to the master telescope layout shown at the bottom of Fig. \ref{fig:Layout_ALL_Prod3}, as detailed in \cite{ICRC2017:Layout}. 
Changing the scaling, each telescope is moved radially so that its new position ($x$, $y$) satisfies $\sqrt{x^2 + y^2} = r \cdot D(r)$, where $r$ is the distance to the centre of the array before the applied transformation and $D(r)$ is the distortion factor, shown in Fig. \ref{fig:distfactor} (top-left).
These transformations change the inter-telescope distance from close to optimal for the low/mid energies to increasingly larger separations for the higher energies.
As an example, the five resulting scaled arrays for one CTA-South layout are shown in Fig. \ref{fig:distfactor}. 
By studying the performance of each simulated scaling, we attempt to find the optimal layout that balances reconstruction quality and event quantity.
At the energy range where the LSTs dominate (below $\sim100$ GeV), the influence of the other telescope types is small, therefore LST spacing optimisation is studied independently and their positions are constant among the five different scalings for both sites.

\begin{figure}[p]
\begin{center}
\begin{minipage}[t]{.5\textwidth}
\vspace{0pt}
\centering
\includegraphics[width=\textwidth]{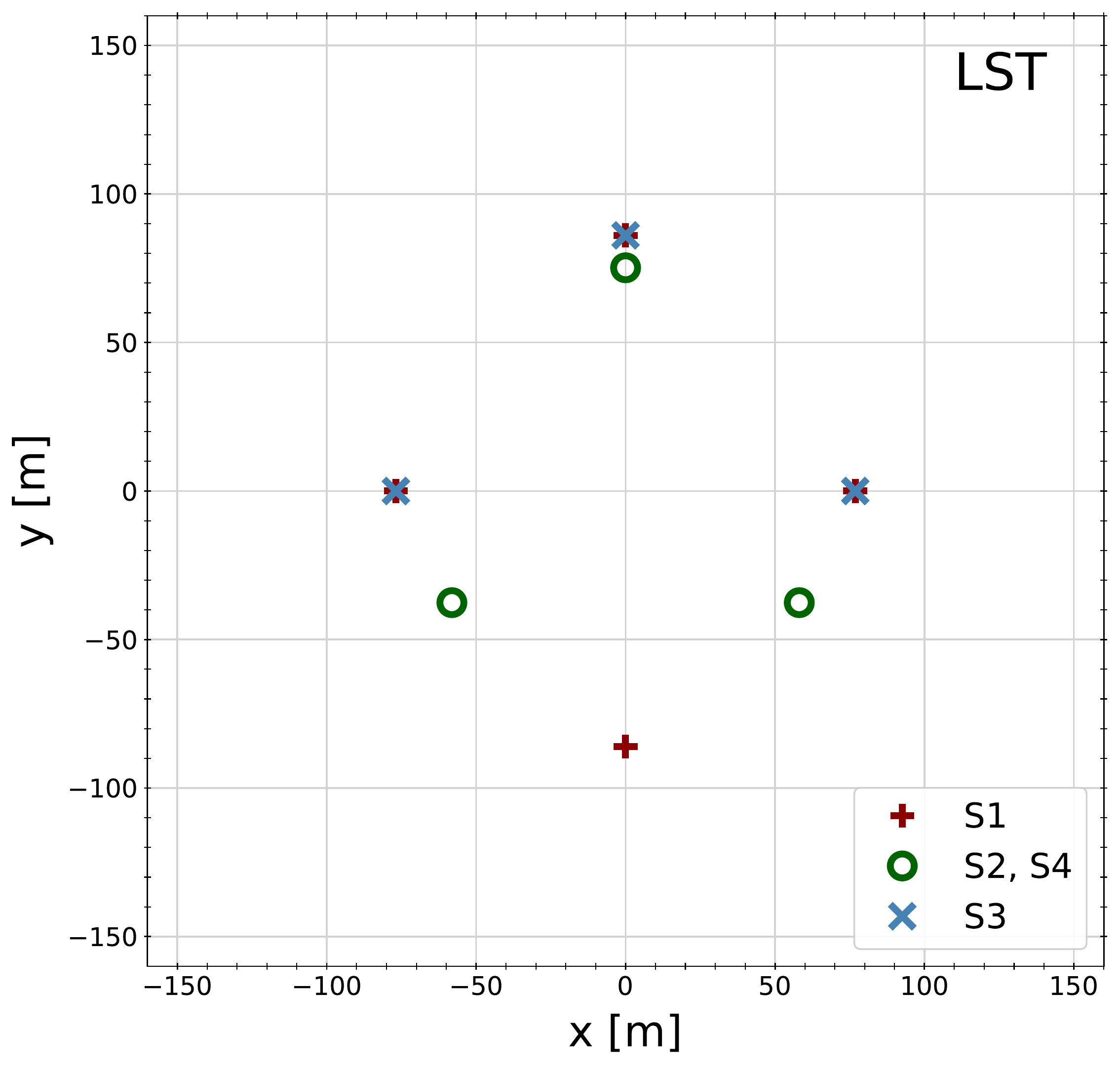}
\end{minipage}%
\begin{minipage}[t]{.5\textwidth}
\vspace{0pt}
\includegraphics[width=\textwidth]{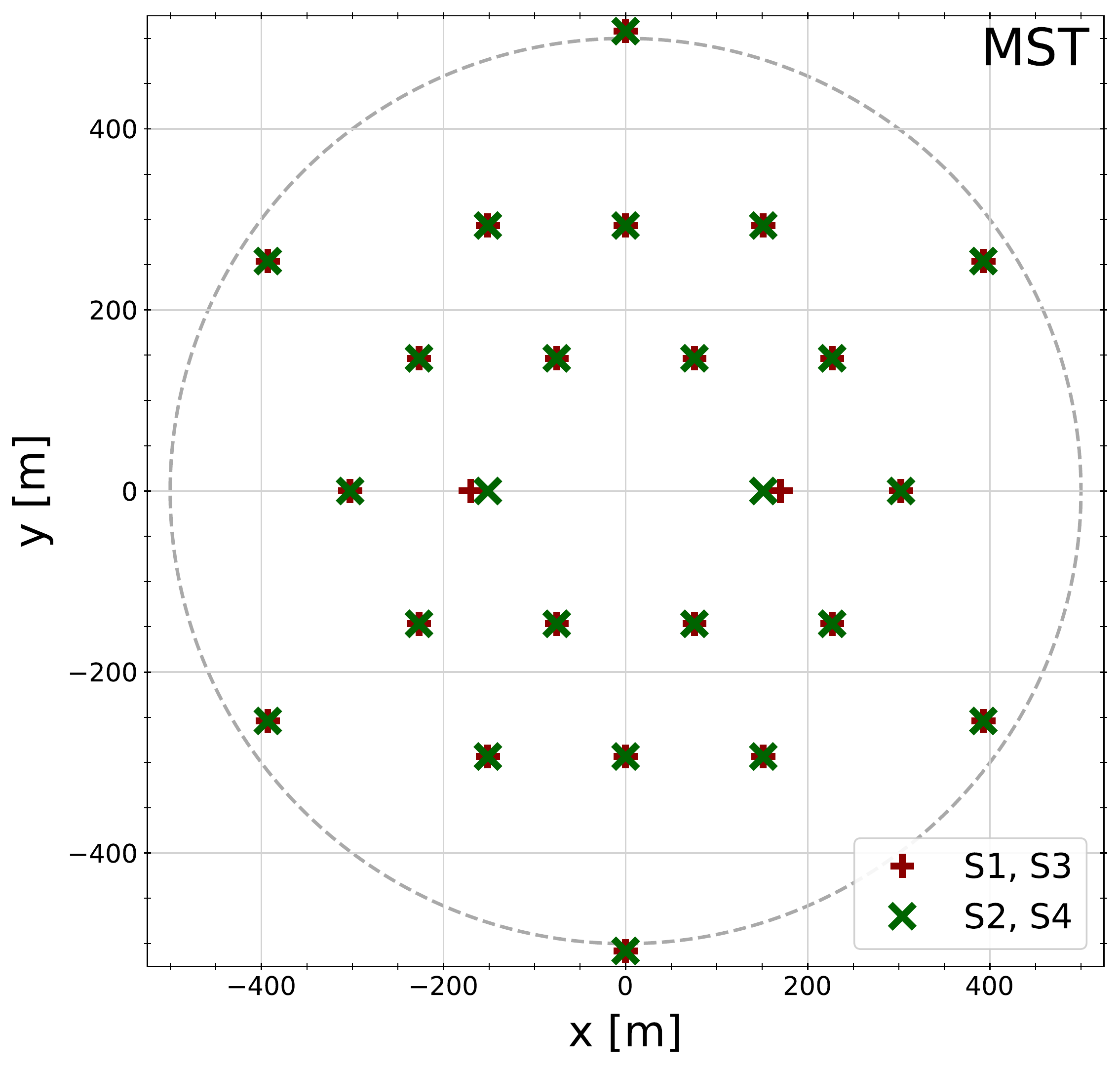}
\end{minipage}%
\\
\begin{minipage}[t]{.5\textwidth}
\vspace{0pt}
\centering
\includegraphics[width=\textwidth]{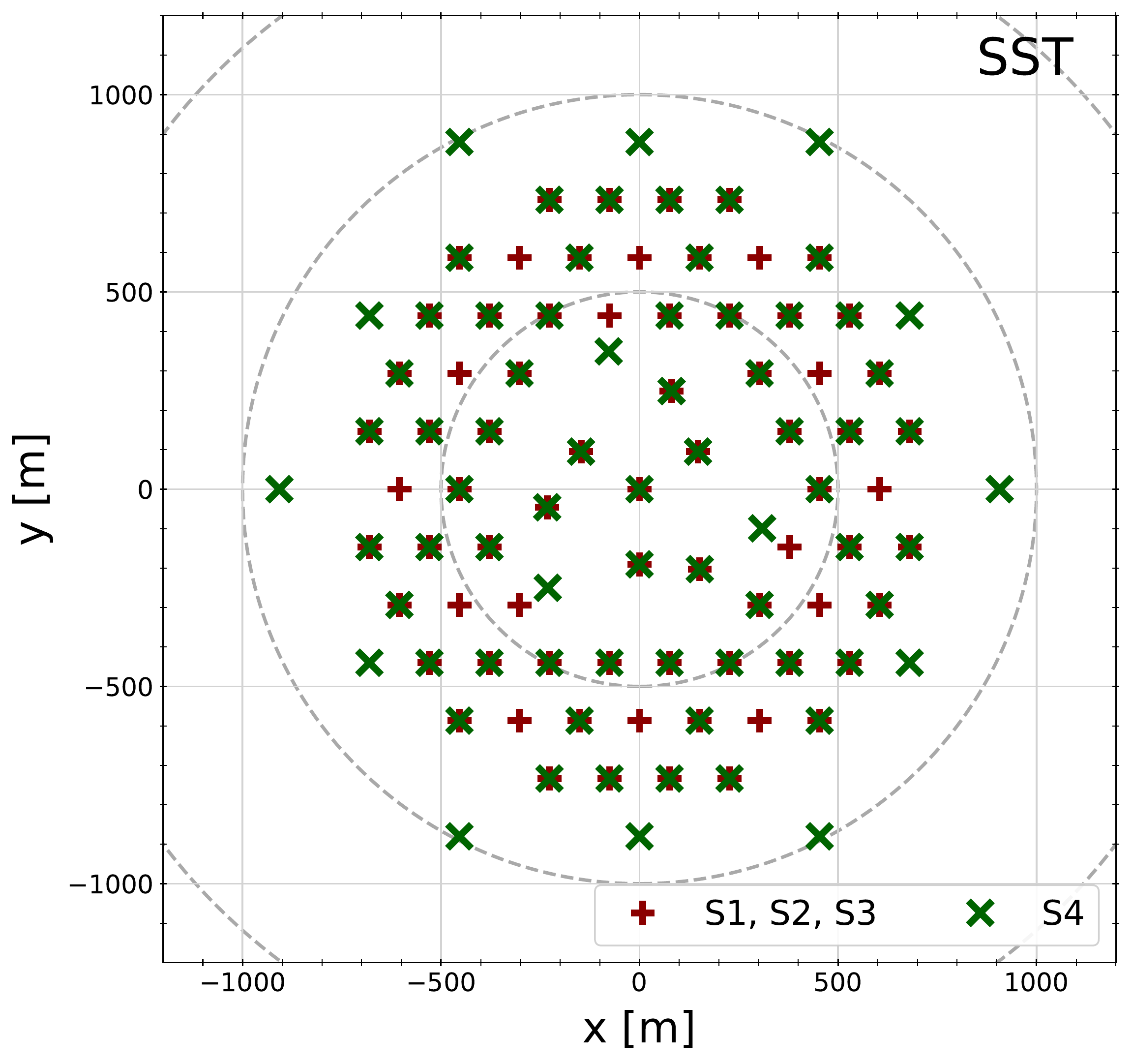}
\end{minipage}%
\begin{minipage}[t]{.5\textwidth}
\vspace{0pt}
\centering
\begin{tabular}{ l c  c c c }
Name  & LST & MST & SST \\\hline
SI-N$_{scaling}$ & 4 & 24 & 72 \\
S1-N$_{scaling}$ & 4 & 24 & 73 \\
S2-N$_{scaling}$ & 3 & 24 & 73 \\
S3-N$_{scaling}$ & 3 & 24 & 73 \\
S4-N$_{scaling}$ & 3 & 24 & 73 \\
S7 & 3 & 24 & 73\\
S8 & 4 & 25 & 70\\
      \end{tabular}
\end{minipage}

\caption{Simulated telescope positions for the different CTA-South array candidates. The positions of each telescope sub-system is shown separately for the arrays ``S1’’ to ``S4’’. The table shows the number of telescopes per type for all layout candidates.}
\label{fig:naming}
\end{center}
\end{figure}

The layout naming convention used throughout the text is the following:
All layout names start with either the letter ``S’’, for CTA-South candidates, or ``N’’, for CTA-North candidates, followed by a number indicating the array variant. When referring to the different scalings of each candidate, an additional number is added after the layout name, e.g. ``S2-3’’ indicates the scaling 3 of the layout ``S2’’. This scheme has two exceptions: the layout ``SI-$N_{scaling}$’’, with an alternative MST distribution shown in Fig. \ref{fig:Layoutisland}, and layouts ``S7’’ and ``S8’’, products of the merging between different scalings, shown in Fig. \ref{fig:3HB9Layout} and discussed in section \ref{sec:souths}. The telescope number and positions of the CTA-South array candidates are shown in Fig. \ref{fig:naming}.

The total number of simulated unique telescope positions adds up to 892 for the southern site and 99 for the northern site. 
At the time the layouts were defined, different alternative designs for the medium and small size telescopes were under consideration and the number of telescopes of each design was not yet fixed. To ensure that the layout resulting from the optimisation does not depend on a certain telescope model, all prototype designs and cameras were simulated, resulting in a total of 3092 simulated telescopes. This way, the performance of each proposed baseline array can be studied for all the different combinations of MST/SST models.

\subsection{Analysis and evaluation criteria}
\label{sec:anchain}

In order to perform the telescope layout optimisation, parameters describing the performance of a given layout need to be defined and maximised. 
As in \cite{Site_Paper}, the primary criteria used in this work to evaluate performance is the differential sensitivity, i.e. the minimum detectable flux from a steady source over a narrow energy range and a fixed observation time. This parameter depends on the collection area, angular resolution and rate of background events, mostly composed by cosmic-ray hadrons and electrons that
survive the gamma-ray selection criteria (cuts). The differential sensitivity is calculated by optimising in each energy bin the cuts on the shower arrival direction, background rejection efficiency and minimum telescope event multiplicity\footnote{The event multiplicity is the number of telescopes simultaneously detecting a shower.}. It is computed by requiring a five standard deviation (5$\sigma$) detection 
significance in each energy bin (equation 17 from \cite{LiMa}, with an off-source to on-source exposure ratio of five, assuming a power-law spectrum of $E^{-2.6}$), and the signal excess to be at least five times the expected systematic uncertainty in the background estimation (1\%), and larger than ten events.

The figure of merit used for the evaluation and comparison of the scientific performance of CTA layouts is called the performance per unit time (PPUT). PPUT is the unweighted geometrical mean of the reference point-source flux sensitivity, $F_\mathrm{sens,ref}$, to the achieved sensitivity, $F_\mathrm{sens}$, over a given energy range with $N$ logarithmically uniform bins (five per decade) in energy:

\begin{equation}
\mathrm{PPUT} =  \left( \prod_{i=1}^{N} \frac{F_\mathrm{sens,ref}(i)}{F_\mathrm{sens}(i)} \right)^{1/N}
\end{equation}

The reference sensitivity was derived from the analysis of previous simulations carried out by the \gls{cta} Consortium, based on initial and conservative assumptions on the telescope parameters (see \cite{APP_CTA_MC}). These reference values, together with other performance requirements (e.g. minimum angular and energy resolutions), constitute the prime goals of the CTA design concept. PPUT may be calculated for the whole CTA-required energy range to estimate the overall performance, i.e. from 20 GeV up to 300 (50) TeV for CTA-South (North), or for energy sub-ranges, to evaluate specific telescope sub-system capabilities. PPUT is defined such that a larger number corresponds to better performance. Statistical uncertainties of all PPUT values, calculated by propagating the differential sensitivity errors associated with the MC event statistics, are below the 3\% level. When comparing PPUT values, these uncertainties are unrealistic given that the performance of all layouts in a given site are calculated from the same set of simulated showers. Statistical uncertainties of PPUT values are therefore not shown in this work.

Except if specified differently, all performance curves and PPUT values shown in this work correspond to a \gls{cta} differential sensitivity to a point-like source in the centre of the FoV with an observation time of 50 hours. The sensitivity of these layouts to sources located at larger angular distances from the centre of the FoV was also evaluated. All telescope layouts presented here were required to comply with a minimum off-axis performance: the radius of the FoV region in which the point-source sensitivity is within a factor two of the one at the centre must be larger than 1$^\circ$ for the LST sub-system (array composed by all and only LSTs) and larger than 3$^\circ$ for the MST and SST sub-systems.

Two fully independent analysis chains, Eventdisplay \cite{evnDisplay} and MARS \cite{MARS} (thoroughly tested by the VERITAS and MAGIC collaborations, respectively), have been used to process the full MC production (at 20$^\circ$ zenith angle) for a large number of telescope configurations for both the Paranal and La Palma sites. In addition, the ImPACT analysis \cite{2014APh....56...26P} was used to produce a cross-check for a small subset of these configurations and the baseline analysis \cite{APP_CTA_MC} was used to validate some results on same-type telescope sub-systems. Eventdisplay, MARS and the methods of the baseline analysis perform classical analyses based on second moment parameterisation of the Cherenkov images \cite{hillas}, with different choice of algorithms for image cleaning, background suppression (Boosted Decision Trees, Random Forest or Lookup tables) and energy reconstruction (Lookup tables or Random Forest). ImPACT is based on a maximum likelihood fit of shower images to pre-generated MC templates, and has proven effective in the analysis of H.E.S.S. data.
In all four cases, background suppression cuts are tuned to achieve the best performance (maximising sensitivity) in each bin of reconstructed energy. See \cite{Site_Paper, APP_CTA_MC} for more details on the analysis. 

\begin{figure}[t]
\begin{center}
\centering\includegraphics[width=0.9\linewidth]{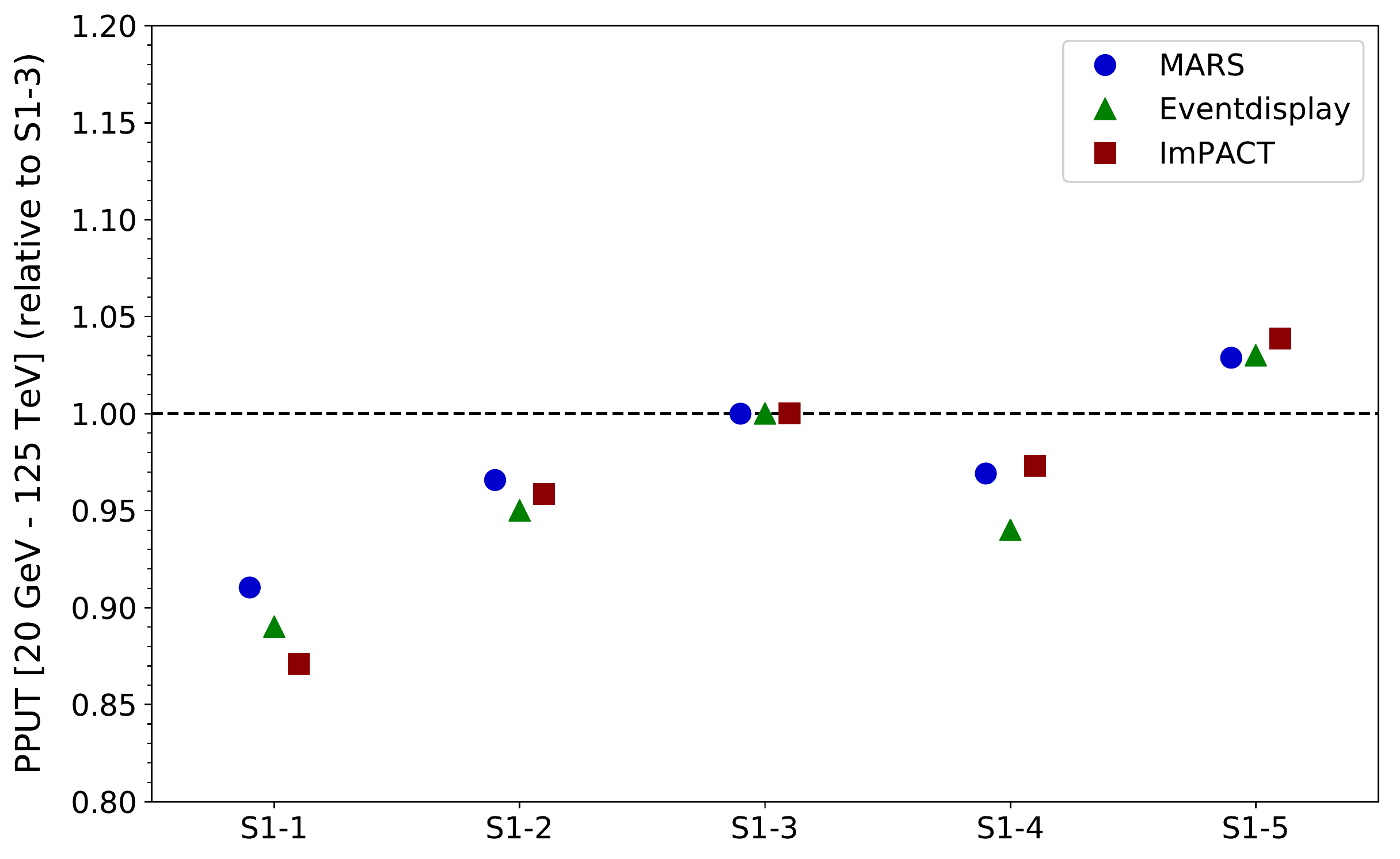}
\caption{Comparison of performance (expressed in terms of PPUT, see text) of a range of simulated array layouts for three different analysis chains, relative to the PPUT value attained by each of them on the ``S1-3’’ layout. The five layouts are presented in Fig. \ref{fig:distfactor}.
The symbols shown in the legend indicate the various analysis chains.}
\label{fig:difanchain}
\end{center}
\end{figure}

Figure \ref{fig:difanchain} shows the PPUT values (between 20 GeV to 125 TeV) of the five scalings simulated for a given CTA-South array candidate, analysed with three of the analysis chains described. The results of the different analyses are, in general, fairly consistent. 
As shown in Fig. \ref{fig:difanchain}, despite their small differences, the conclusion on the optimal layout is the same regardless of the choice of analysis package. 

\subsection{Telescope Configurations}

The third large-scale MC production was simulated using the most realistic and detailed modelling of all \gls{cta} telescopes and camera types available.
Given that the prototype telescopes were in the development stage
at the time of the production (summer 2015), some telescope and camera parameters used within these models may be different from the final ones.
These differences are expected to have a small effect on single-telescope performance, so all conclusions inferred from this study will still be valid, as long as the \gls{cta}-proposed telescopes do not undergo major design changes.

SC-MSTs were excluded from this study due to technical limitations. The limited available memory during computation did not allow the production of sufficient event statistics for their performance evaluation. Given the relatively similar mirror area and FoV of DC-MSTs and SC-MSTs, it is unlikely that the replacement of some DC-MSTs with SC-MSTs in the proposed layouts would result in a sub-optimal array layout.

\begin{figure}[t]
\begin{center}
\centering\includegraphics[width=0.9\linewidth]{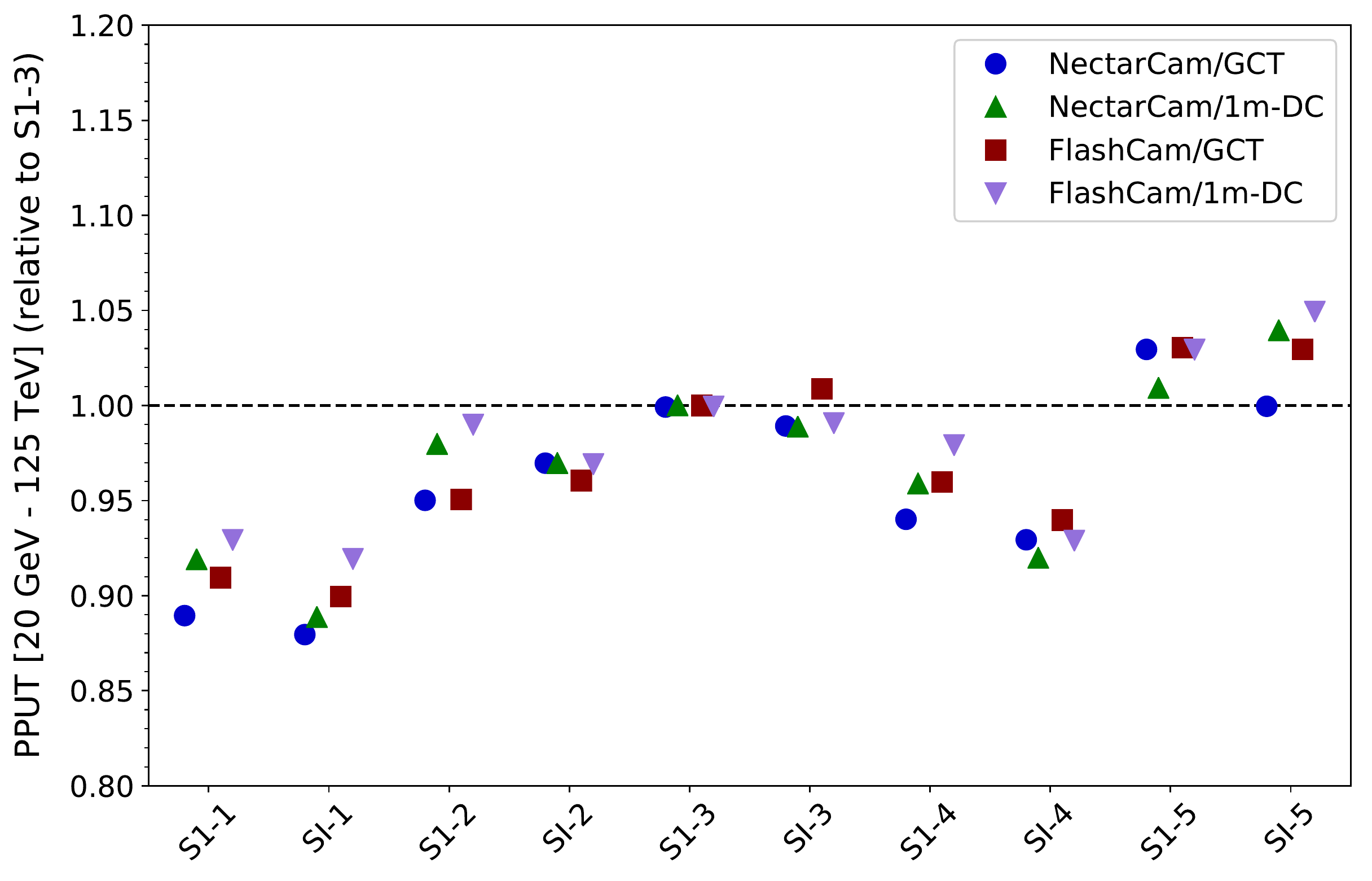}
\caption{Comparison of performance (expressed in terms of PPUT, see text) of a range of simulated CTA-South array layouts for different combinations of telescope model configurations, each relative to the ``S1-3’’ layout. The different ``S1’’ layout scalings are pictured in Fig. \ref{fig:distfactor}, while the ``SI’’ layouts are described in section \ref{sec:MSTSSTpat}.
The symbols shown in the legend indicate the various telescope configurations.
}
\label{fig:PPUT_NFDG}
\end{center}
\end{figure}

As the final configuration of \gls{cta} telescope types is not known at this point (e.g. how many SSTs of each design will be constructed), the analysis always considers arrays of a single MST and SST design. All possible combinations between the two DC-MST cameras and the three SST models have been studied to ensure that the layout choice does not depend on specific telescope configurations. Figure \ref{fig:PPUT_NFDG} shows as an example the PPUT values of some CTA-South arrays using different combinations of telescope models: NectarCam/GCT, NectarCam/SST-1M, FlashCam/GCT, and FlashCam/SST-1M. The relative differences of the PPUT values between the different configurations for a given array layout are below 5\% and clearly show the same trend upon changes of the array layout and scaling.

\section{Layout Optimisation}
\label{sec:conlayout}

The final numbers of telescopes of each type is now fixed for both hemispheres, defined as the most cost-effective solution to maximise \gls{cta} performance over the key scientific cases \cite{ScienceCTA}.
The number of telescopes that the baseline arrays will be composed of are 4/25/70 LST/MST/SST for CTA-South and 4/15 LST/MST for CTA-North. With the number of telescopes fixed, the layout optimisation was performed following these considerations (in approximate order of priority):

\begin{enumerate}
    \item[C1.] Full system performance requirements.
    \item[C2.] Telescope sub-system performance requirements (e.g. MST-only array performance).
    \item[C3.] Topographical constraints of the selected sites.
    \item[C4.] Shadowing between neighbouring telescopes (i.e. telescopes structure intersecting the FoV of other telescopes during large zenith angle observations).
    \item[C5.] Performance of partially-operating arrays (e.g. resulting from telescope staging or downtime).
    \item[C6.] Impact on the ease of calibration and the likely magnitude of systematic effects.
\end{enumerate}

For C1, the main optimisation parameter is the differential sensitivity of the full array, while simultaneously ensuring that the energy resolution, the angular resolution and the FoV requirements are still met. C2 ensures that the system works in a close-to-optimal fashion also when operated as individual (LST, MST or SST) sub-systems. C3 is critical for the northern site (La Palma), but was not needed for the southern site, where no significant constraints are expected. C4 sets a minimum telescope spacing for pairs of each telescope size combination. If possible, without moving significantly away from the optimum performance for the baseline, point C5 was addressed by ensuring that partially completed systems are still close to optimal. In the case of the LSTs, of which only four telescopes will be installed on each site, the effect of telescope downtime was taken into consideration due to the expected occasional maintenance of one of these telescopes. For MSTs and SSTs, a few missing telescopes due to maintenance is not expected to significantly affect the performance. Finally, point C6 was addressed by requiring some overlap between different telescope sub-systems even when the array is partially completed.

\subsection{LST optimal separation}
\label{subsec:lsts}

\begin{figure}[hpt]
\begin{center}
\includegraphics[width=0.45\linewidth]{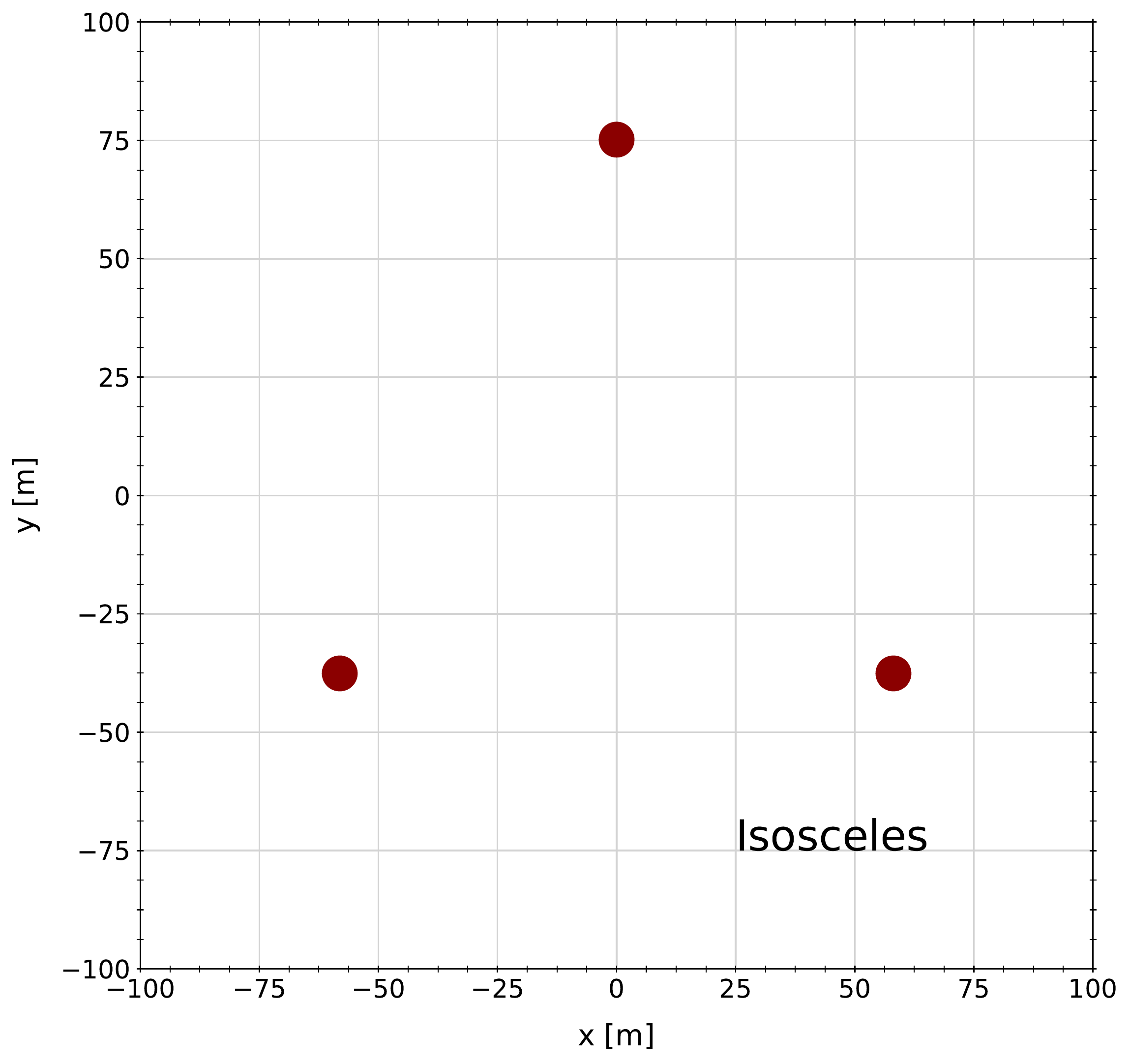}
\includegraphics[width=0.45\linewidth]{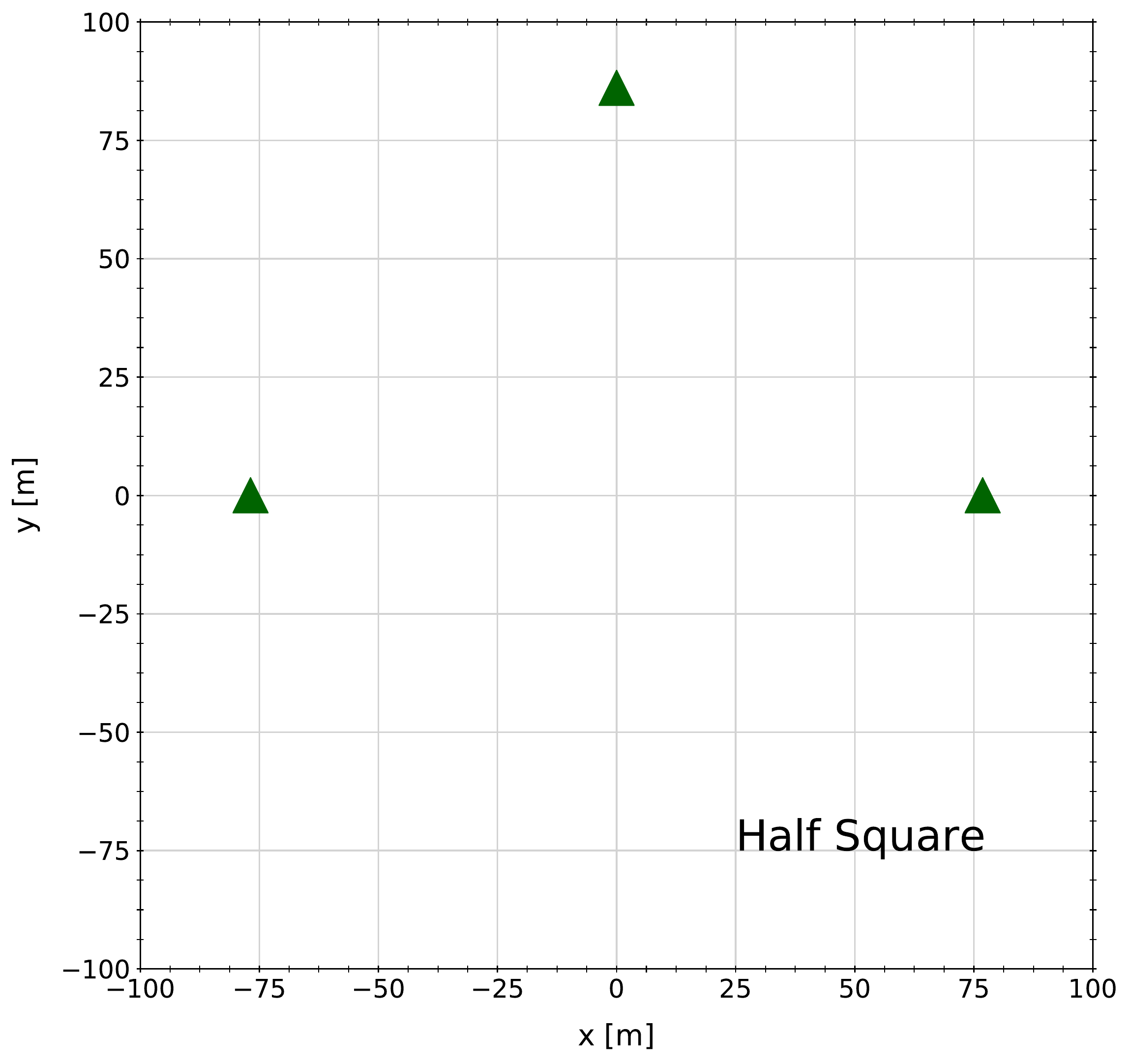}
\includegraphics[width=0.9\linewidth]{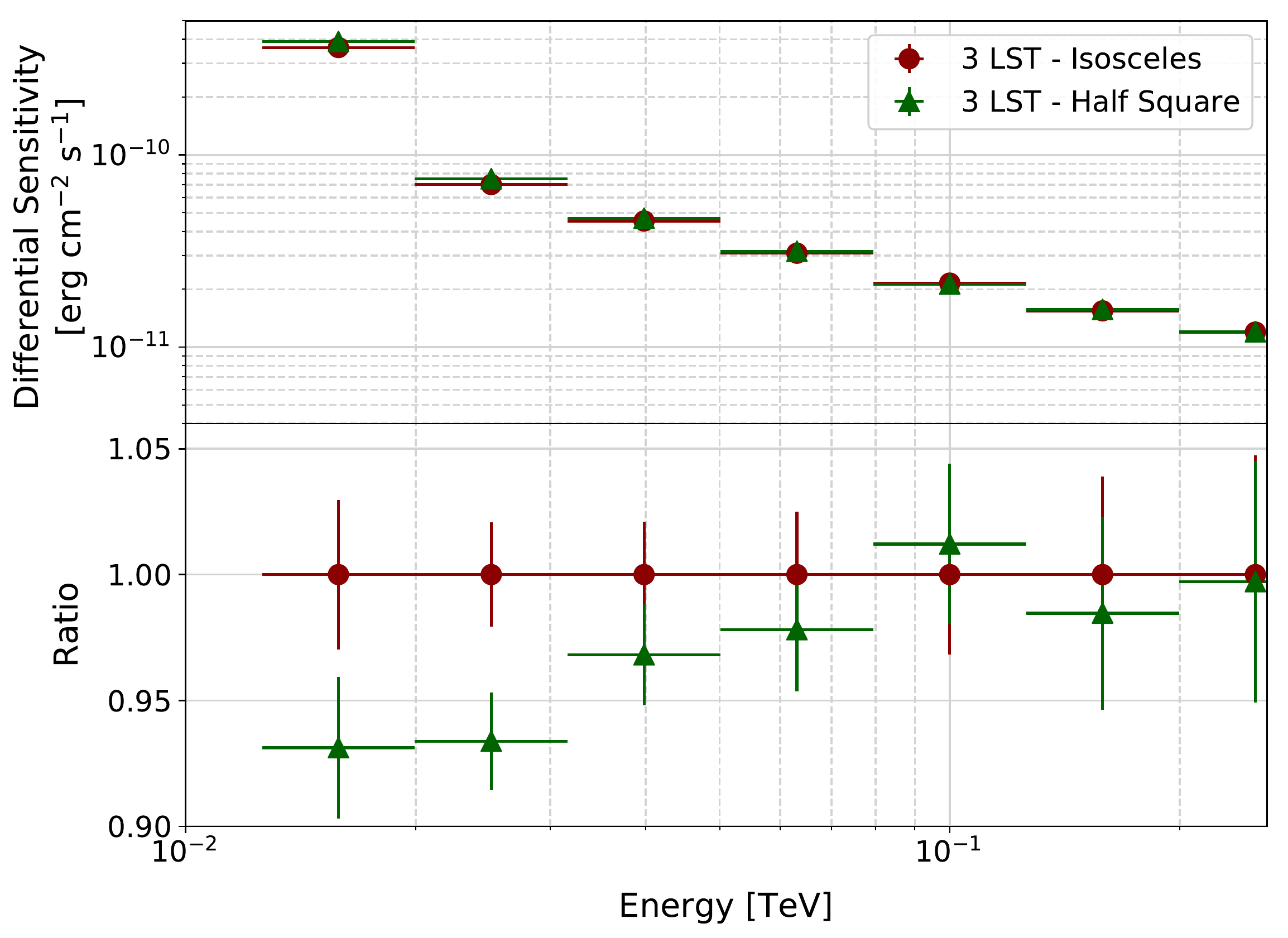}
\caption{Differential sensitivity and differential sensitivity ratio as a function of energy 
for two configurations of three LSTs with equal area (\textit{bottom}): arranged as half a square of 115 m on a side (\textit{top right}) or an isosceles triangle with two 127 m sides (close to equilateral, \textit{top left}). The layouts are slightly stretched in the north-south direction and compressed in the east-west direction, as explained in section \ref{SimTelLay}. The ratio is calculated so that higher values correspond to better sensitivity.}
\label{fig:3LST}
\end{center}
\end{figure}

Below $\sim100$~GeV the LSTs will dominate CTA performance, as these will be the only telescopes with enough reflecting surface to detect the faint low-energy showers.
For this reason, the layout of the MST and SST positions have no strong impact in this energy range, therefore their spacing optimisation can be studied independently. These 
showers are generally triggered within impact distances\footnote{The impact distance is the between the telescope location and the shower axis.} below 150~m, i.e. similar to the light pool radius of about 120 m \cite{Site_Paper}. As the light-pool size increases with the energy of the primary particle, the optimal LST spacing is expected to be smaller than for MSTs or SSTs.

The optimal shape of the LST sub-system in the shower-plane projection is expected to be a square for four LSTs and an equilateral triangle for three LSTs. This is confirmed in Figure \ref{fig:3LST}, which shows the low-energy differential sensitivity of a three LST layout with an isosceles shape, close to equilateral, compared to a three LST layout with a half-square shape.

The optimisation of the LST layout beyond these considerations is thus a question of separation only. At too-short separations, the projected lever arm in the stereoscopic shower reconstruction is too small for most events while at too-large separations too few showers are detected simultaneously by three or four LSTs (required for an optimal cosmic-ray background rejection).

As described in section \ref{sec:mcprod}, the second large-scale MC production assessed CTA performance over a wide range of site candidates. Realistic values of the altitude and geomagnetic field strength at each site were used in the shower simulation \cite{Site_Paper}. Nine different LST positions were included at each site, allowing the analysis of several equivalent layouts (e.g. pairs of two LSTs) with different inter-telescope distances. Archival simulation sets for the following \gls{cta} site candidates were available for this analysis (see \cite{Site_Paper} for details on each site): Aar (near Aus, Namibia) at 1640~m altitude, two sites at Leoncito (Argentina) at 1650 and 2660 m, and SAC (San Antonio de los Cobres, Argentina) at 3600 m altitude. To test the array performance at lower altitudes, an additional hypothetical Aar site was simulated at 500 m altitude. For the SAC site candidate, at whose altitude the Cherenkov light pool is significantly smaller, an additional set of simulations were performed with the telescope spacing reduced by a factor of 0.84, allowing us to test a larger number of telescope distances.

\begin{figure}[t]
\begin{center}
\centering\includegraphics[width=0.9\linewidth]{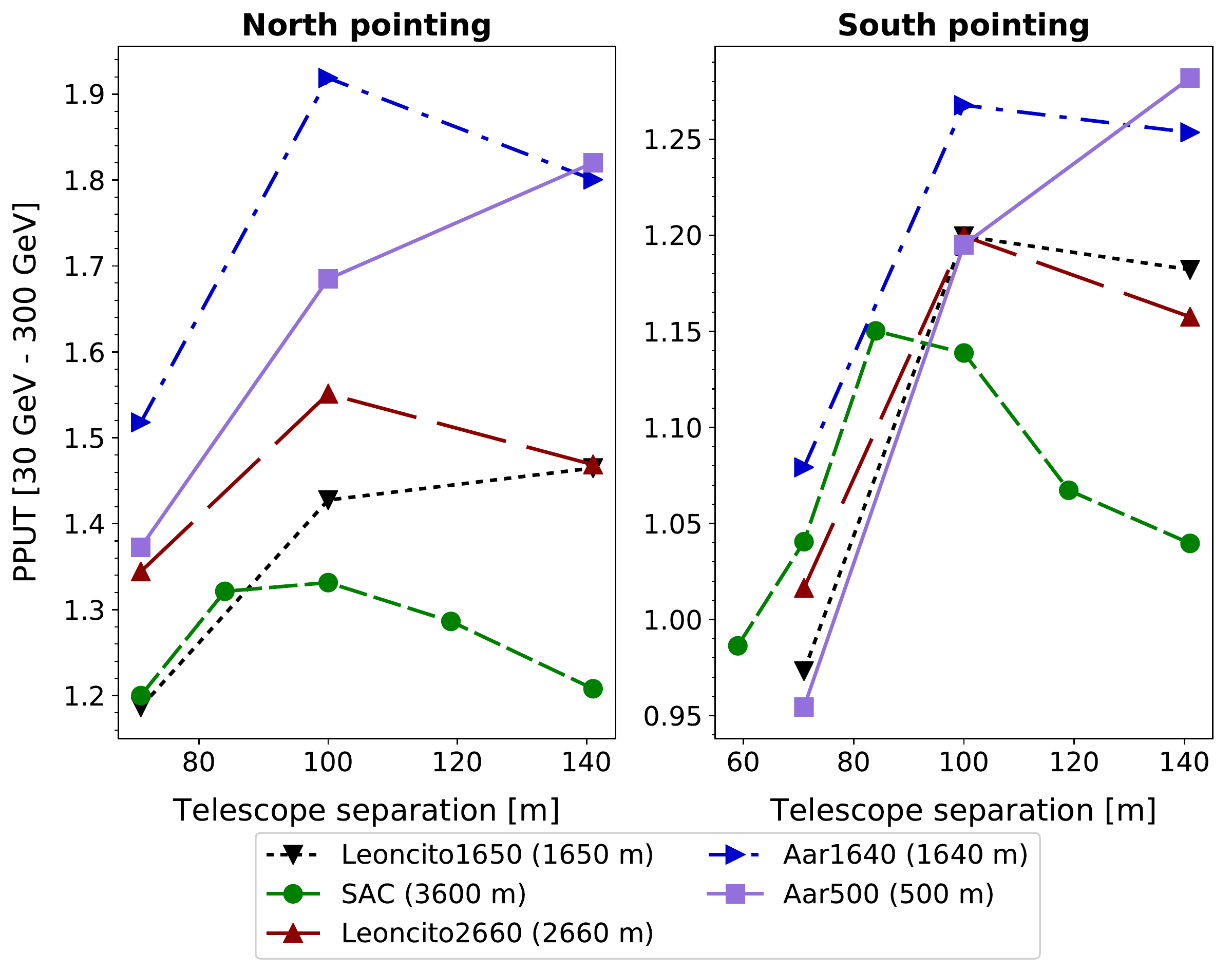}
\caption{Performance (expressed in terms of PPUT, see text) of LST squared layouts of different sizes located at different CTA-South candidate sites (\textit{left}: observations towards north, \textit{right}: observations towards south), 
in the energy range 30 to 300 GeV, using the baseline analysis described in \cite{APP_CTA_MC}.}
\label{fig:lstseparationsquare}
\end{center}
\end{figure}

For a layout of four LSTs in a square shape, side distances of 71, 100, and 141~m (plus 59, 84, and 119 m only for SAC) were available. Figure~\ref{fig:lstseparationsquare} shows the dependence of the LST sub-system performance versus telescope separation for all the studied sites.
For the Paranal site, with an altitude and geomagnetic field falling between the two simulated Leoncito sites shown in Fig. \ref{fig:lstseparationsquare}, a  separation of about 100 m (square side length) is favoured.

\begin{figure}[t]
\begin{center}
\centering\includegraphics[width=0.9\linewidth]{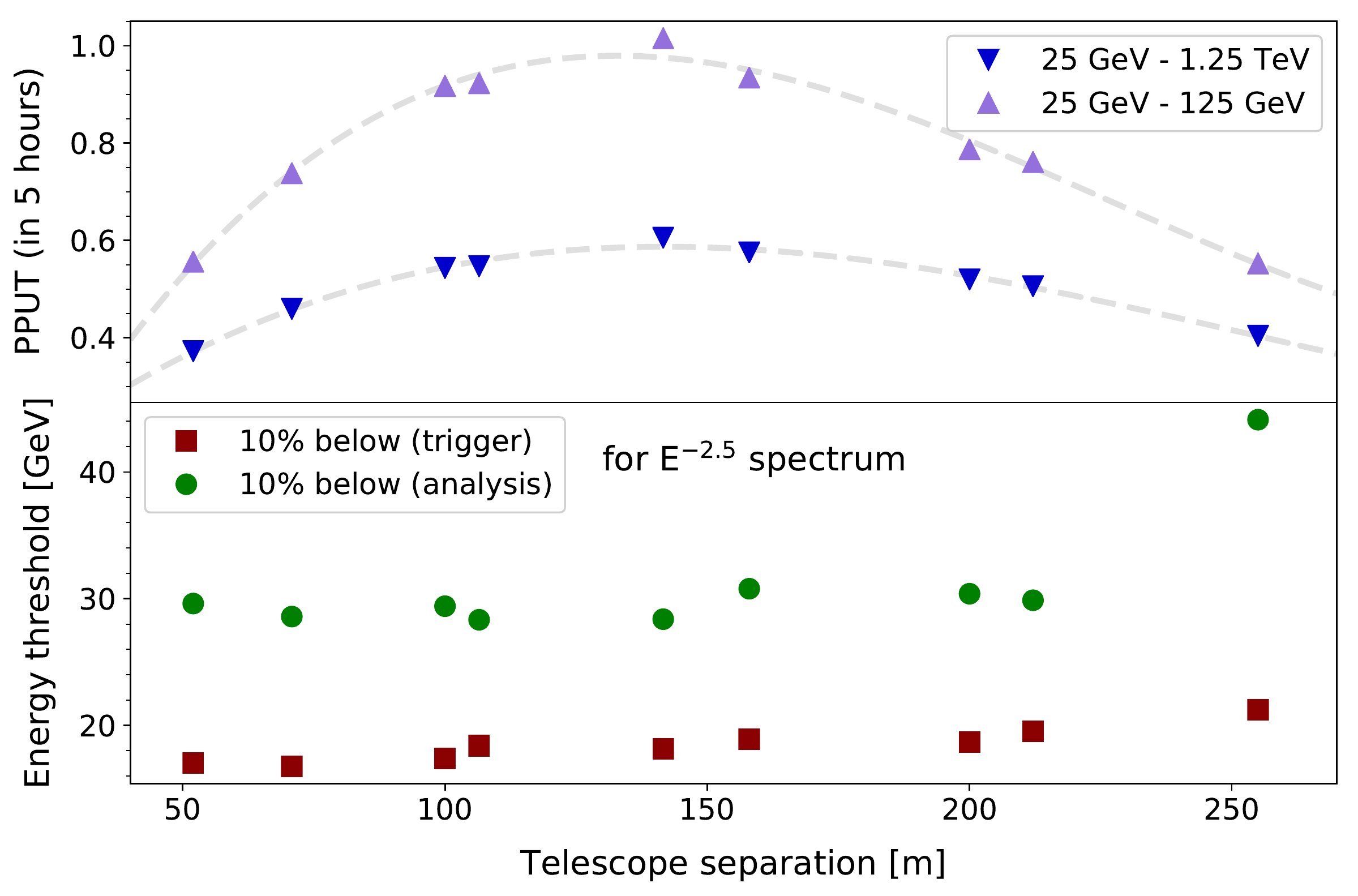}
\caption{Performance (expressed in terms of PPUT, see text) and energy threshold of pairs of LSTs as a function of their separation. PPUT values are calculated from the average of the Aar and the two Leoncito site candidates (with an average altitude close to that of the Paranal site) and are also averaged over observations pointing towards north and south. The upper panel shows PPUT values in the energy ranges of 25 GeV to 125 GeV and 25 GeV to 1.25 TeV; the lower panel shows the calculated energy threshold by using the true energy value that leaves 10\% of the events below the cut value (after either the trigger or the analysis) \cite{Colin2009}. The performance is derived from the baseline analysis described in \cite{APP_CTA_MC}.}.
\label{fig:lstseparation}
\end{center}
\end{figure}

For the case of LST pairs, there were nine different distances available between 58 to 255 m. As shown in Figure \ref{fig:lstseparation}, a rather flat optimum is found at 130 m, with close-to-optimum performance for separations ranging from about 100 m up to 150 m, with no significant change in energy threshold over this range. The optimum separation over the whole LST energy range (more relevant for observation with the LST sub-system only) is not significantly larger than for just the lowest energies (relevant for observations with the full array).

Taking all these results into account, a squared layout of four LSTs with an optimised side distance of 115 m to 120 m would provide both full-system and sub-system optimal performance. In order to make sure the rest of the listed considerations, such as geological constrains for the La Palma site or improved staging scenarios for Paranal, are complied with, minor modifications were needed to be applied to these positions. As shown in Figure \ref{fig:3LST}, such minor modifications of the LST layout are expected to affect the performance at only the few percent level.

\subsection{MST and SST patterns}\label{sec:MSTSSTpat}

\begin{figure}[p]
\begin{center}
\includegraphics[width=0.9\linewidth]{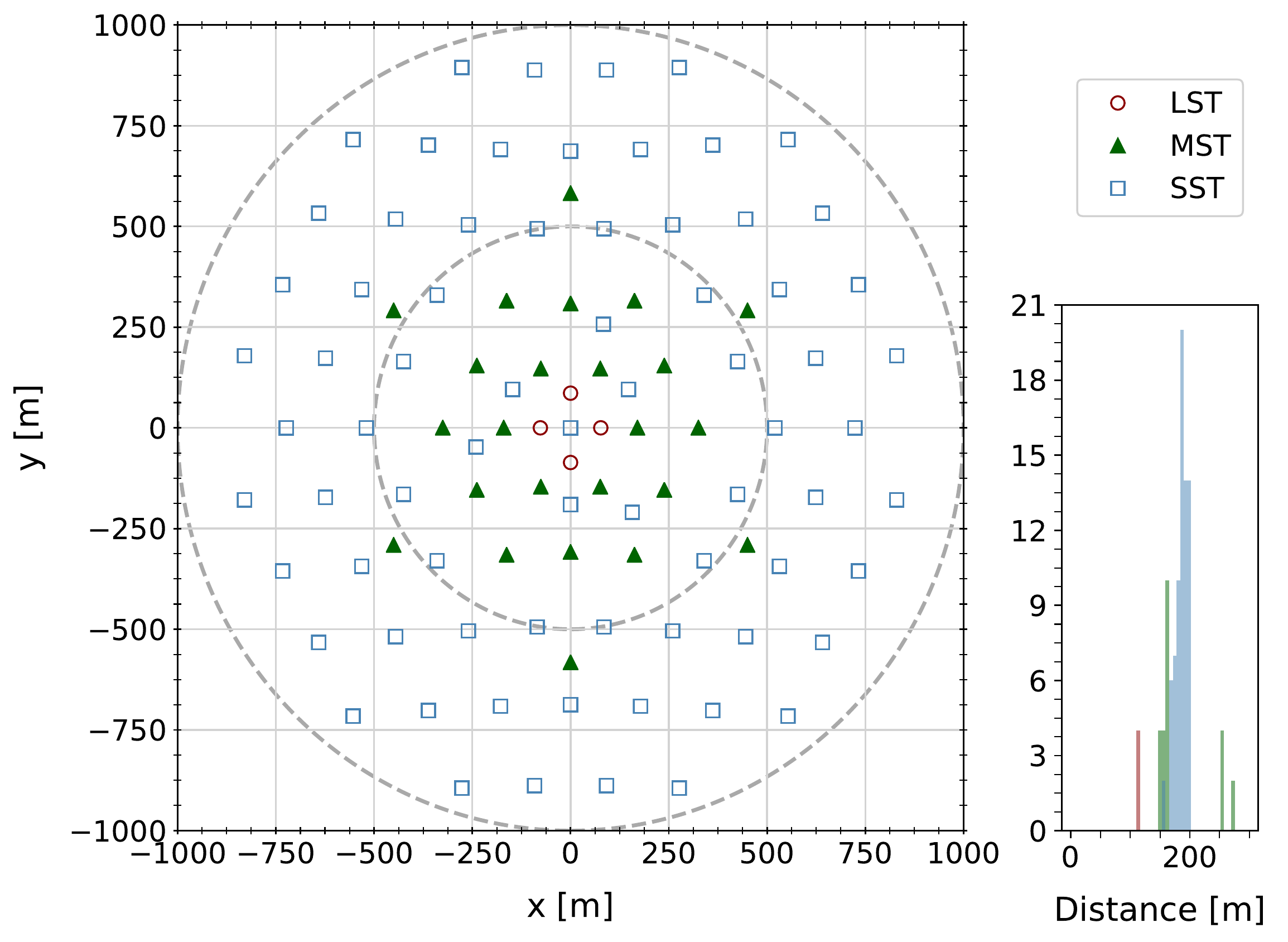}
\includegraphics[width=0.9\linewidth]{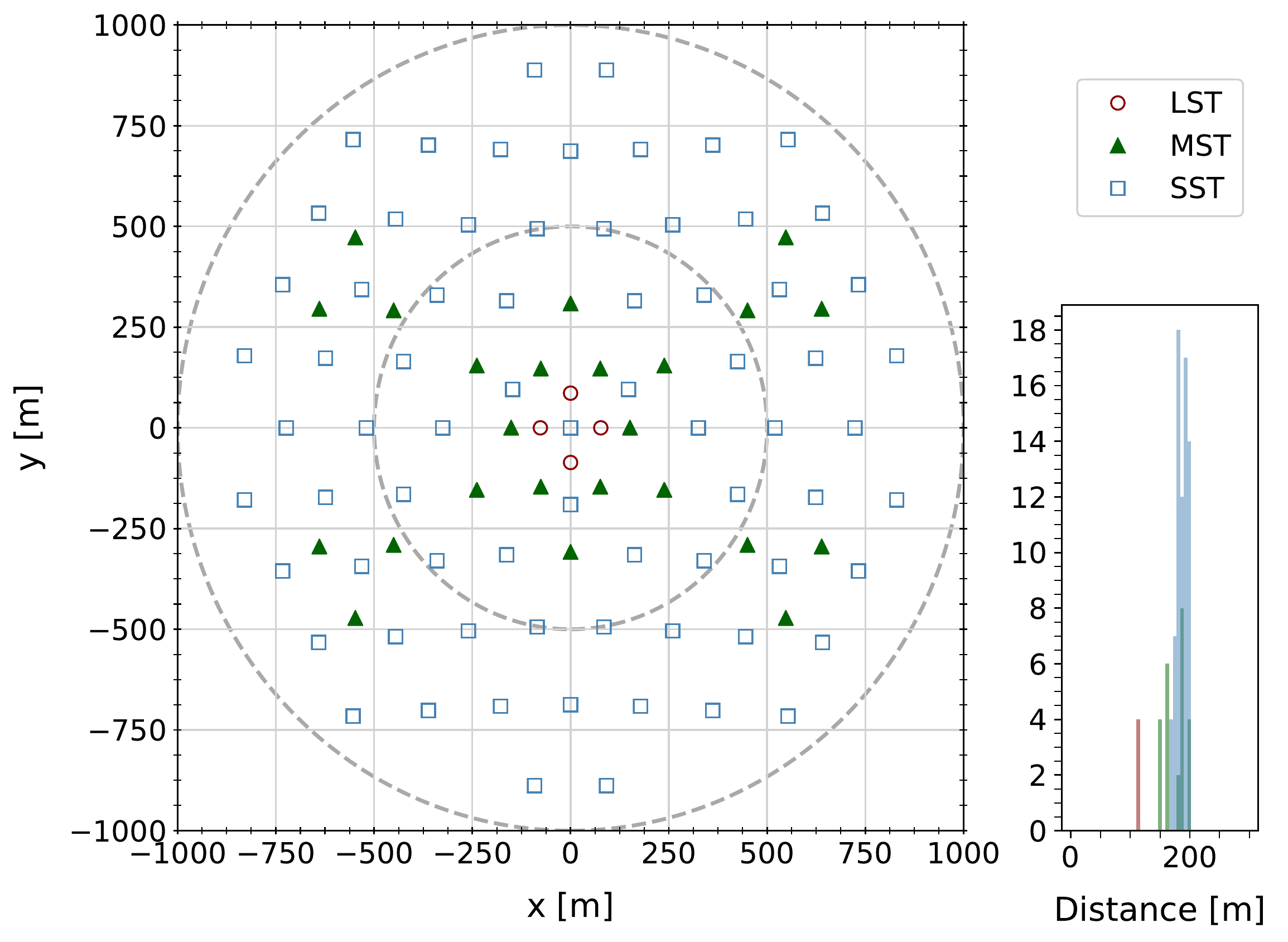}
\caption{Layouts with different MST patterns: ``S1’’ (\textit{top}), with a strictly hexagonal pattern and ``SI’’ (\textit{bottom}), with four islands and a hexagonal core. The LST positioning in the two cases is the same, while the SSTs have been rearranged. Both layouts correspond to their scaling 2 variation. The distance of each telescope to its nearest neighbour of the same type is shown on the right.}
\label{fig:Layoutisland}
\end{center}
\end{figure}

\begin{figure}[t]
\begin{center}
\centering\includegraphics[width=0.99\linewidth]{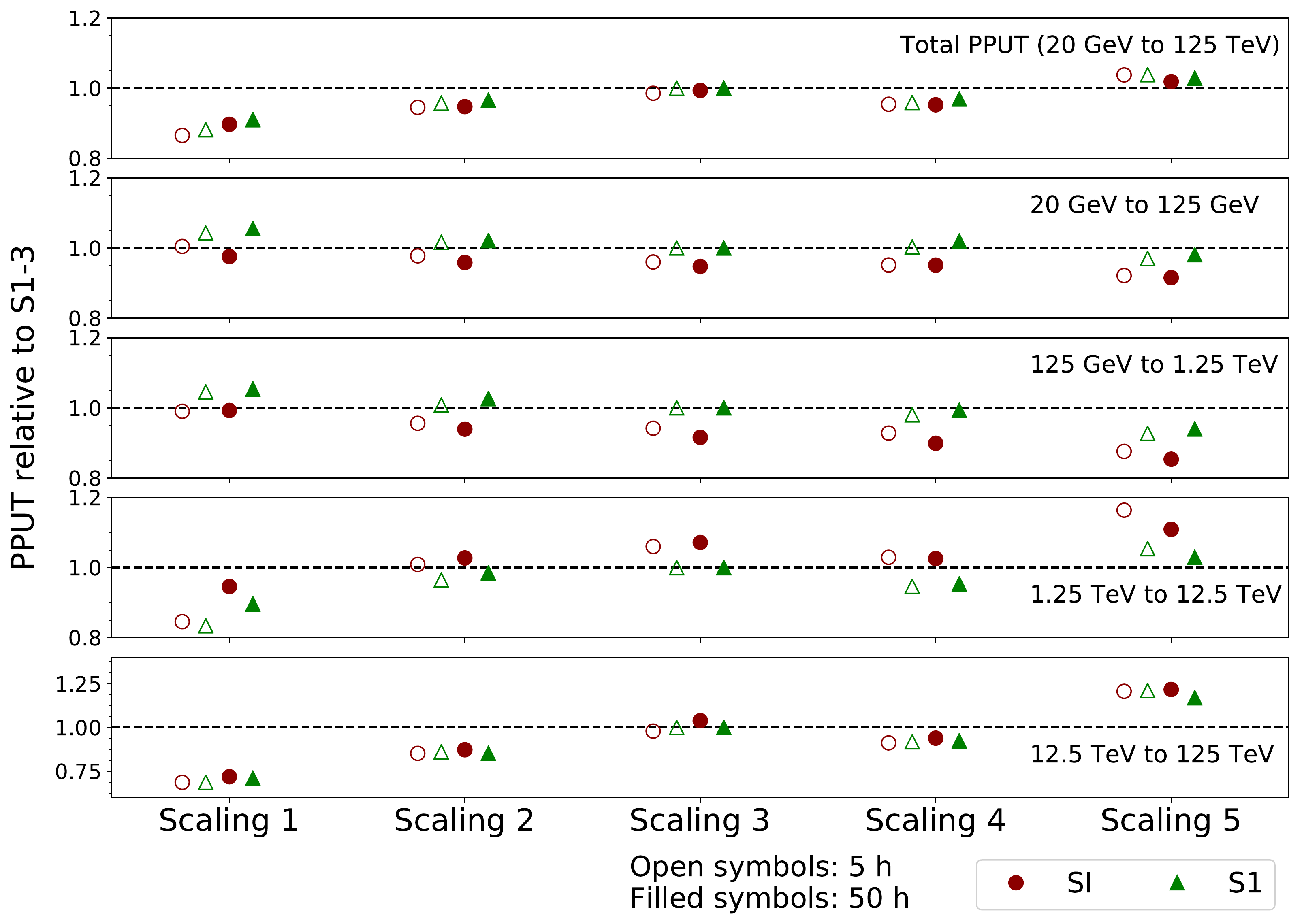}
\caption{Relative PPUT values for different energy ranges for the layout with a hexagonal MST pattern (``S1’’) and a layout with an MST pattern presenting four islands (``SI’’), both for the southern site, relative to ``S1-3’’. Open and filled symbols correspond to observation times of 5 h and 50 h, respectively.}
\label{fig:PPUTisland}
\end{center}
\end{figure}

As introduced in section \ref{sec:mcprod}, the master layout of simulated telescopes used in this work is based on a hexagonal layout to enhance the statistics of showers simultaneously detected by at least three telescopes \cite{Colin_2009}. From this layout, two different MST patterns were studied: a hexagonal one (as in ``S1’’, top of figure \ref{fig:Layoutisland}) and one presenting an inner hexagonal core with fewer telescopes and four surrounding islands of three MSTs each (as in ``SI’’, bottom of figure \ref{fig:Layoutisland}). Because of the repositioning of MSTs, some SSTs have been moved in order to provide uniform coverage. The positions of the LSTs are shared between the two layouts.

As shown in Fig. \ref{fig:PPUTisland}, the two layouts provide comparable overall sensitivity over the whole energy range (20 GeV to 125 TeV). Over the low and medium energy ranges (20 GeV to 1.25 TeV) the hexagonal pattern is preferred, given the higher number of MSTs simultaneously used to reconstruct these contained showers (i.e. showers whose light pools are fully contained inside the area covered by CTA telescopes). Between 1.25 TeV and 12.5 TeV, the island pattern provides better performance due to the improved reconstruction of high-energy showers triggering telescopes near the edge of the array. This improvement fades above 12 TeV, for energies dominated by the SST sub-system. The hexagonal MST pattern was chosen as the preferred option given its improved performance over a wider energy range. Two different observation times were tested in this comparison, 5 and 50 hours, to make sure that the inferred conclusions are not dependent on the observation time.

\section{Southern site baseline array}\label{sec:souths}

\begin{figure}[t]
\begin{center}
\centering\includegraphics[width=0.99\linewidth]{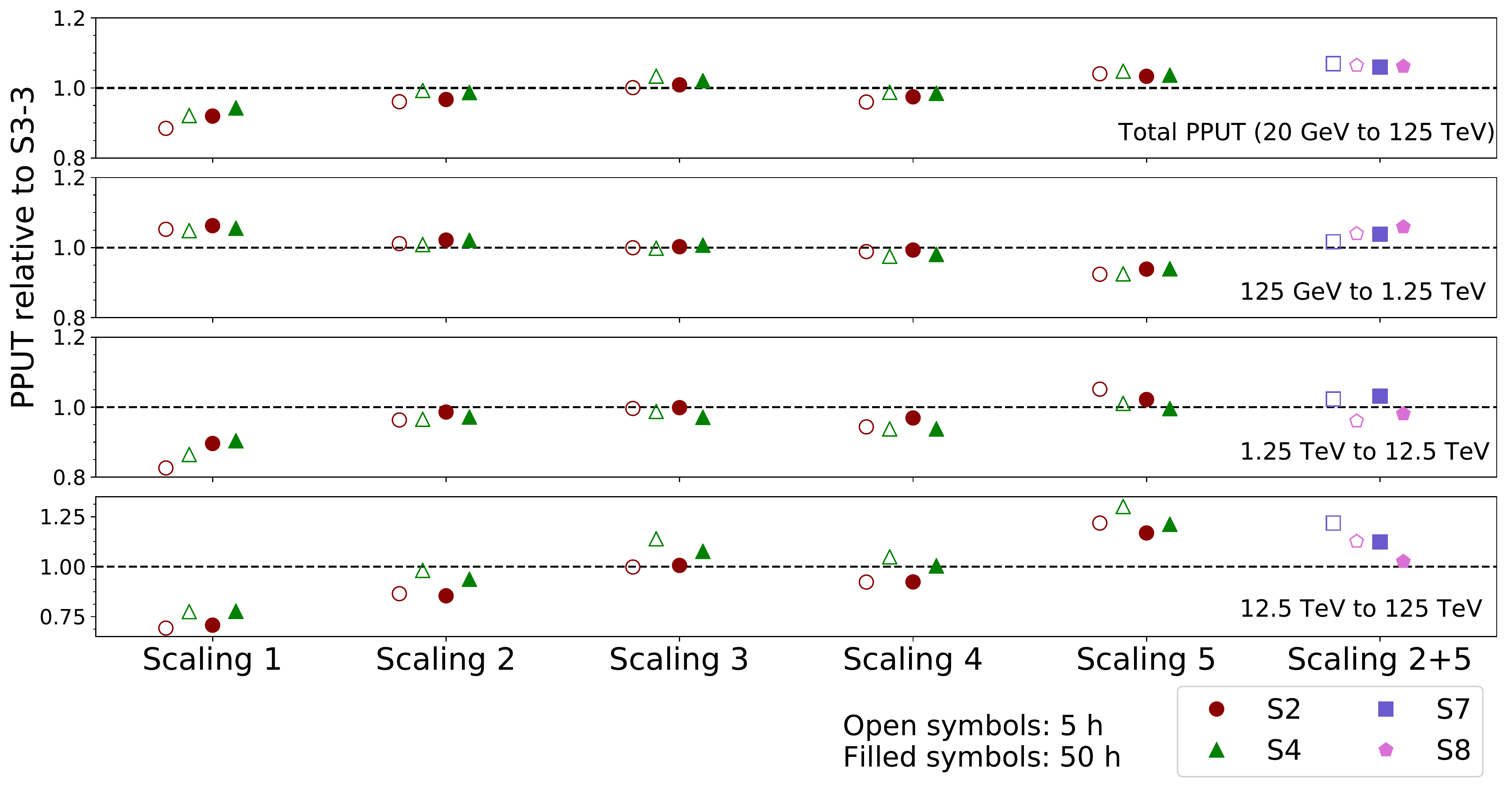}
\caption{Relative PPUT values for different energy ranges for several CTA-South layout candidates, relative to ``S3-3’’. The resulting PPUT values obtained by combining the MST layout with moderate radial scaling (2) and the SST layout with strong scaling (5) are shown labelled as ``scaling 2+5’’.}
\label{fig:PPUTscalings}
\end{center}
\end{figure}

The PPUT values for six different energy ranges were calculated for three different CTA-South layout candidates (``S2’’ and ``S4’’, calculated with respect to ``S3’’) and their five different radial scalings. As shown in Fig. \ref{fig:PPUTscalings}, more compact arrays improve performance below $\sim1$ TeV, but have poorer performance compared to
arrays with larger scalings at higher energies.  
Taking these results into account, a new layout is defined combining the MST layout with moderate radial scaling (2) and the SST layout with strong scaling (5), labelled as ``S7’’. As shown in Fig. \ref{fig:PPUTscalings}, it is the layout with best overall performance, outperforming most alternatives in every energy range.

However, minor modifications are still necessary to be applied to ``S7’’ for two important reasons: 1) it includes slightly different numbers of telescopes with respect to the defined baseline (4 LSTs, 25 MSTs and 70 SSTs) and 2) the distribution of the SSTs is sub-optimal for independent sub-system operation and complicates cross-calibration.
The proposed baseline layout for \gls{cta}-South is therefore a slightly modified version of ``S7’’, named ``S8’’ (both shown in Fig. \ref{fig:3HB9Layout}). The performed modifications are 
discussed below:

\begin{figure}[t]
\begin{center}
\includegraphics[width=0.49\linewidth]{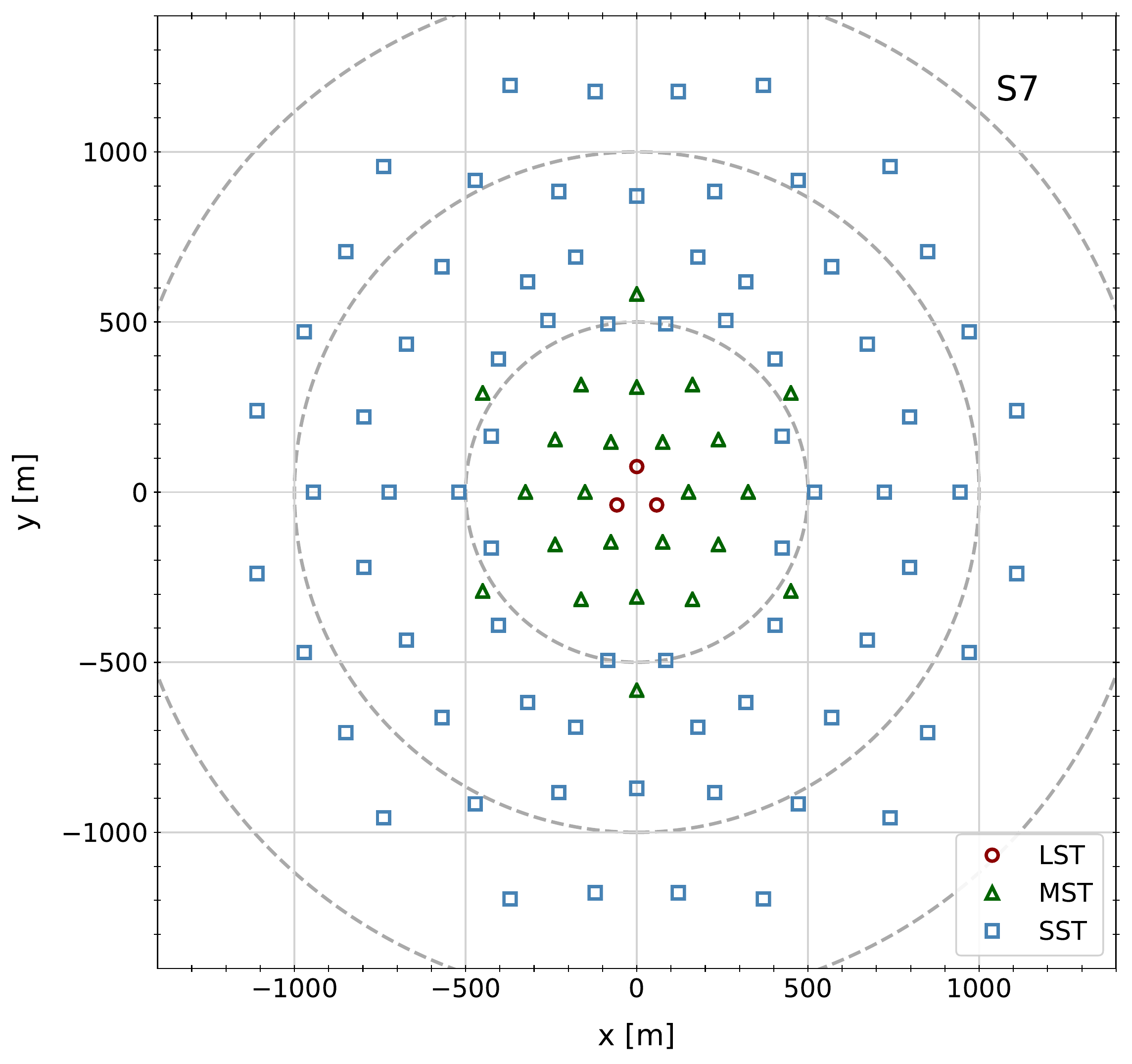}
\includegraphics[width=0.49\linewidth]{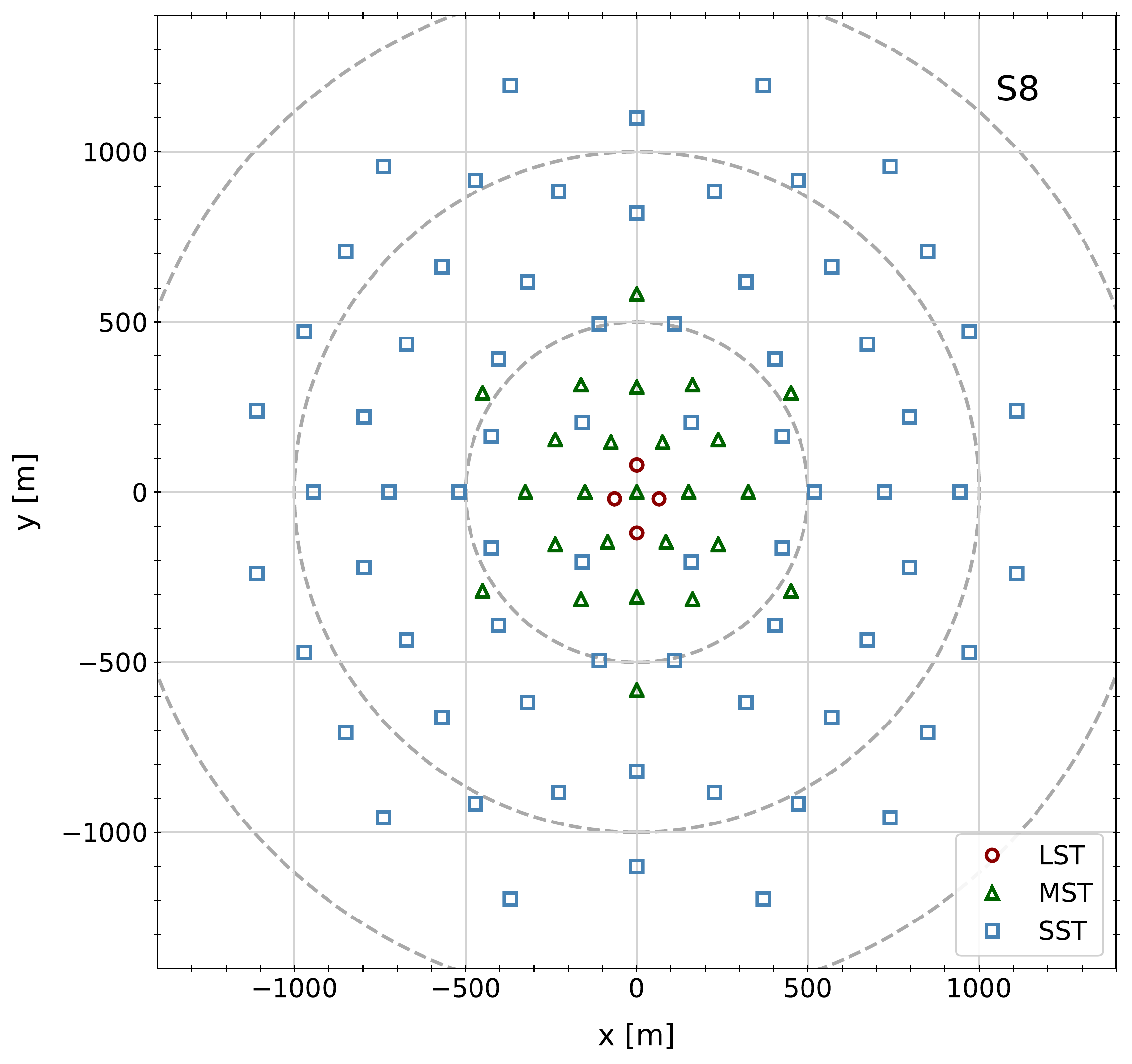}
\caption{The best performing layouts from Fig. \ref{fig:PPUTscalings}: ``S7’’, on the left, and the proposed baseline layout for the southern site, ``S8’’, on the right.}
\label{fig:3HB9Layout}
\end{center}
\end{figure}

\begin{itemize}
    \item The LST layout is rather independent of the optimisation of the system as a whole.
    The proposed four LST layout is an intermediate step between a square and a double-equilateral triangle, with the advantage that it performs significantly better than a square for a three LST stage, without significant degradation of the full system performance. This compromise also works better than the double-equilateral triangle configuration for the situation where one of the east-west pair of telescopes is unavailable (e.g. due to maintenance activities). The east-west pair of telescopes represents the best option for a two LST stage-1\footnote{The east-west telescope pair provides better stereoscopic reconstruction while pointing north/south, the preferred sky directions in which sources culminate.}, and therefore the chosen telescope separation is close to optimal for a two-telescope system (as shown in Sec. \ref{subsec:lsts}).
    \item The MST layout for the proposed array is identical to ``S7’’ except for the addition of a central MST. The central MST is particularly useful for MST sub-system operation, surveying performance and LST-MST cross-calibration.
    \item The SST positions are modified from ``S7’’ by removing four telescopes (``S7’’ has 74 SSTs) and smoothing their distribution. Four SSTs are moved within the boundary of the dense MST array to enhance the SST-only sub-system performance, to provide better MST-SST cross-calibration and to smooth the performance transition between the MST-dominated to 
the SST-dominated energy range. After fixing the four inner telescopes and the outer boundary edge of the layout (so that the highest energy performance is not affected), the spacing of the remaining telescopes is adjusted to minimise the inter-telescope distance.
\end{itemize}

As mentioned in section \ref{sec:mcprod}, some telescope positions within ``S8’’ were not available and needed to be added to the third large-scale MC production.
This extension was necessary to confirm that these modifications were not strongly affecting performance. As shown in Fig. \ref{fig:PPUTscalings}, the overall PPUT of ``S8’’ matches the one attained by ``S7’’. Even if the performance above $\sim1$ TeV is slightly affected by subtracting four SSTs, ``S8’’ outperforms most layout alternatives, while taking into account all considerations listed in section \ref{sec:conlayout}. For these reasons, ``S8’’ is the final telescope layout proposed as the baseline for the CTA southern site.

\section{Northern site baseline array}\label{sec:norths}

\begin{figure}[t]
\begin{center}
\centering\includegraphics[width=0.9\linewidth]{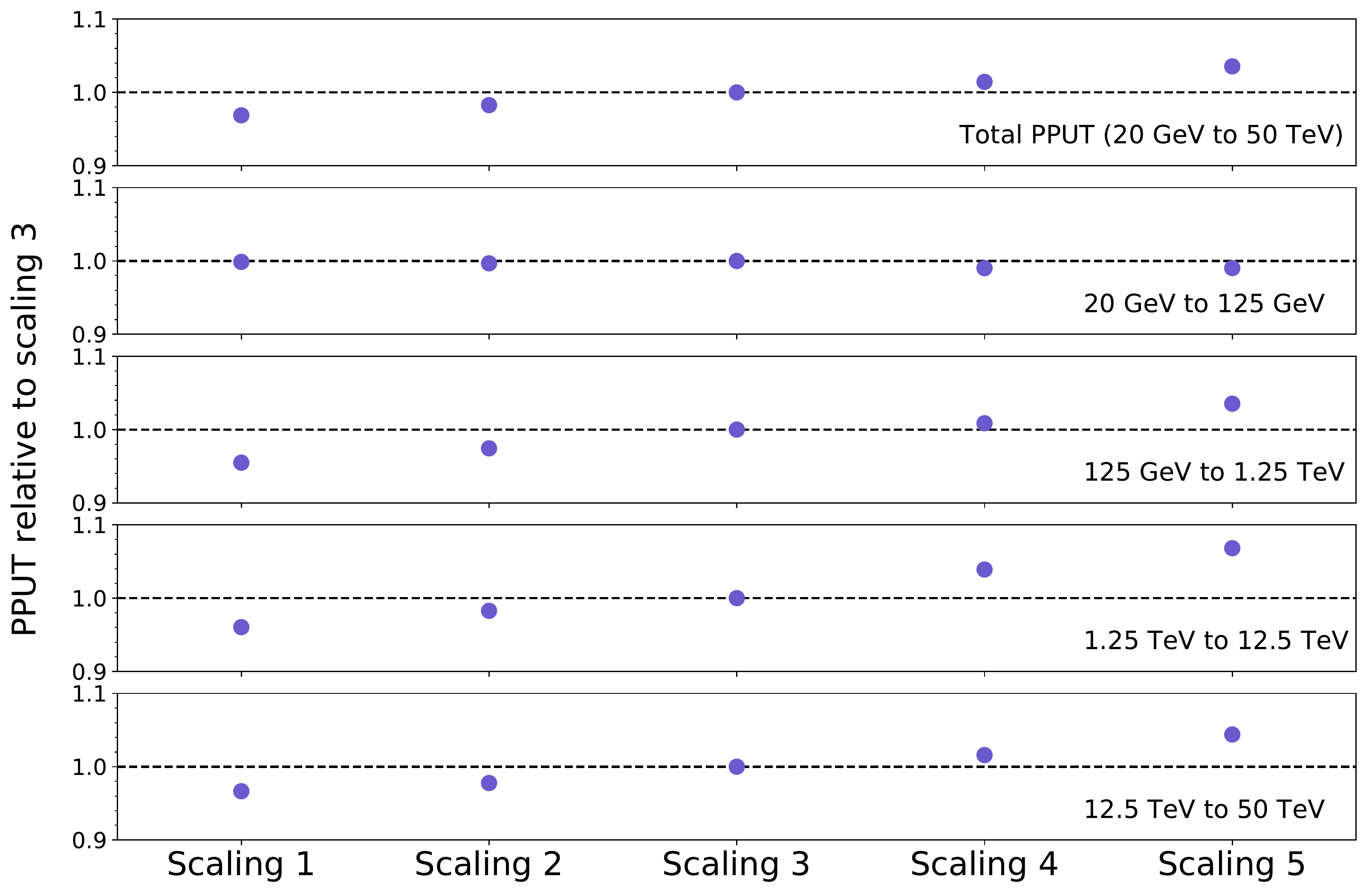}
\caption{Relative PPUT values for the different scalings of the proposed layout for the northern site, all shown in Fig. \ref{fig:Layout_ALL_Prod3}, relative to the scaling 3.}
\label{fig:LaPalmaPPUT}
\end{center}
\end{figure}

As discussed in section \ref{sec:mcprod}, the available telescope positions of the CTA-North layout were mainly constrained by site topography, buildings and roads. As Figure \ref{fig:LaPalmaPPUT} illustrates, the best overall performance from the simulated layouts is achieved by the widest MST spacing considered. This large spacing does not have an impact on the low energy performance while guaranteeing the best sensitivity at higher energies. An even wider spacing, while possible for some of the telescopes, is forbidden by the logistical constraints of the site.

\begin{figure}[p]
\begin{center}
\includegraphics[width=0.49\textwidth]{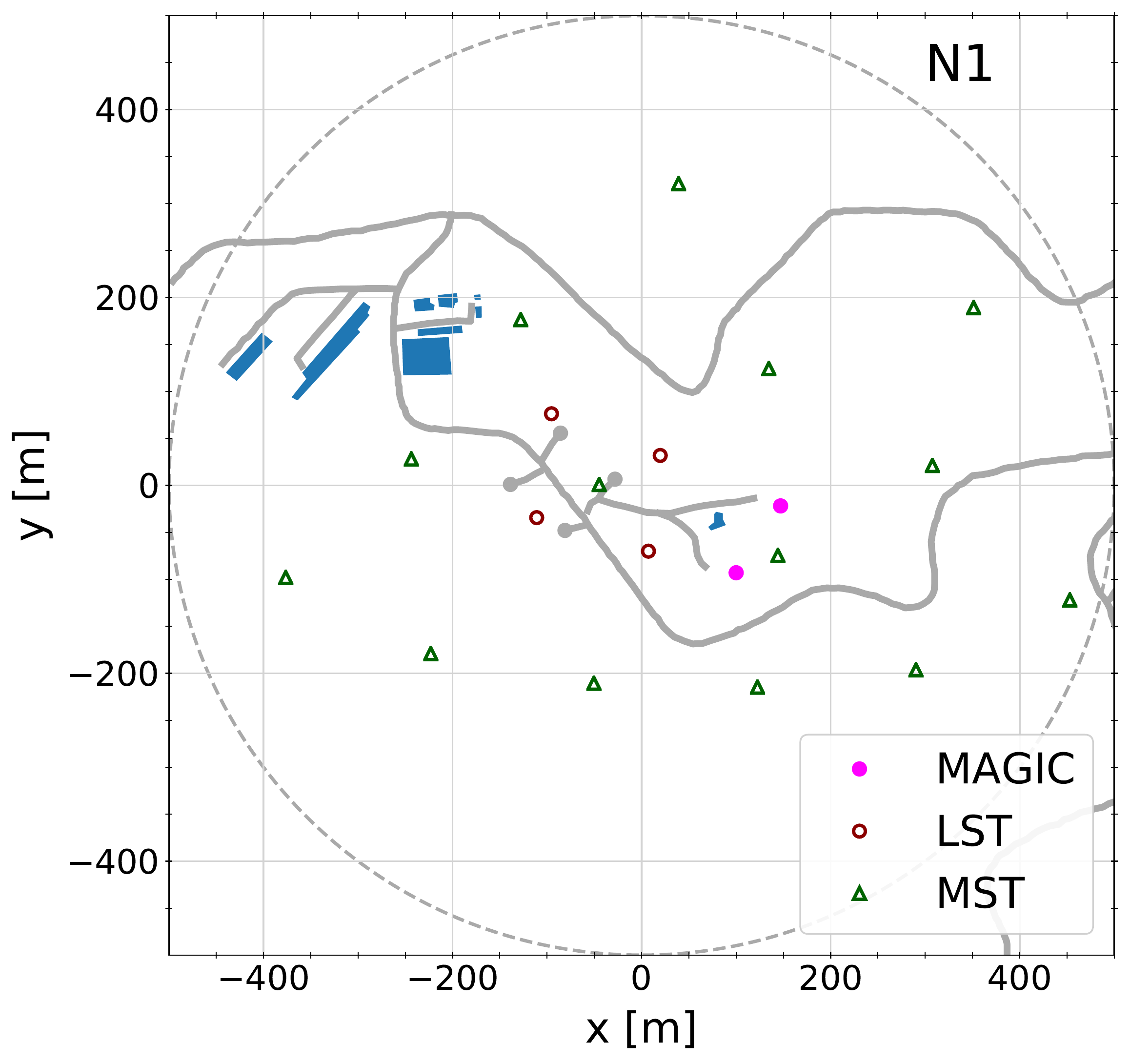}
\includegraphics[width=0.49\textwidth]{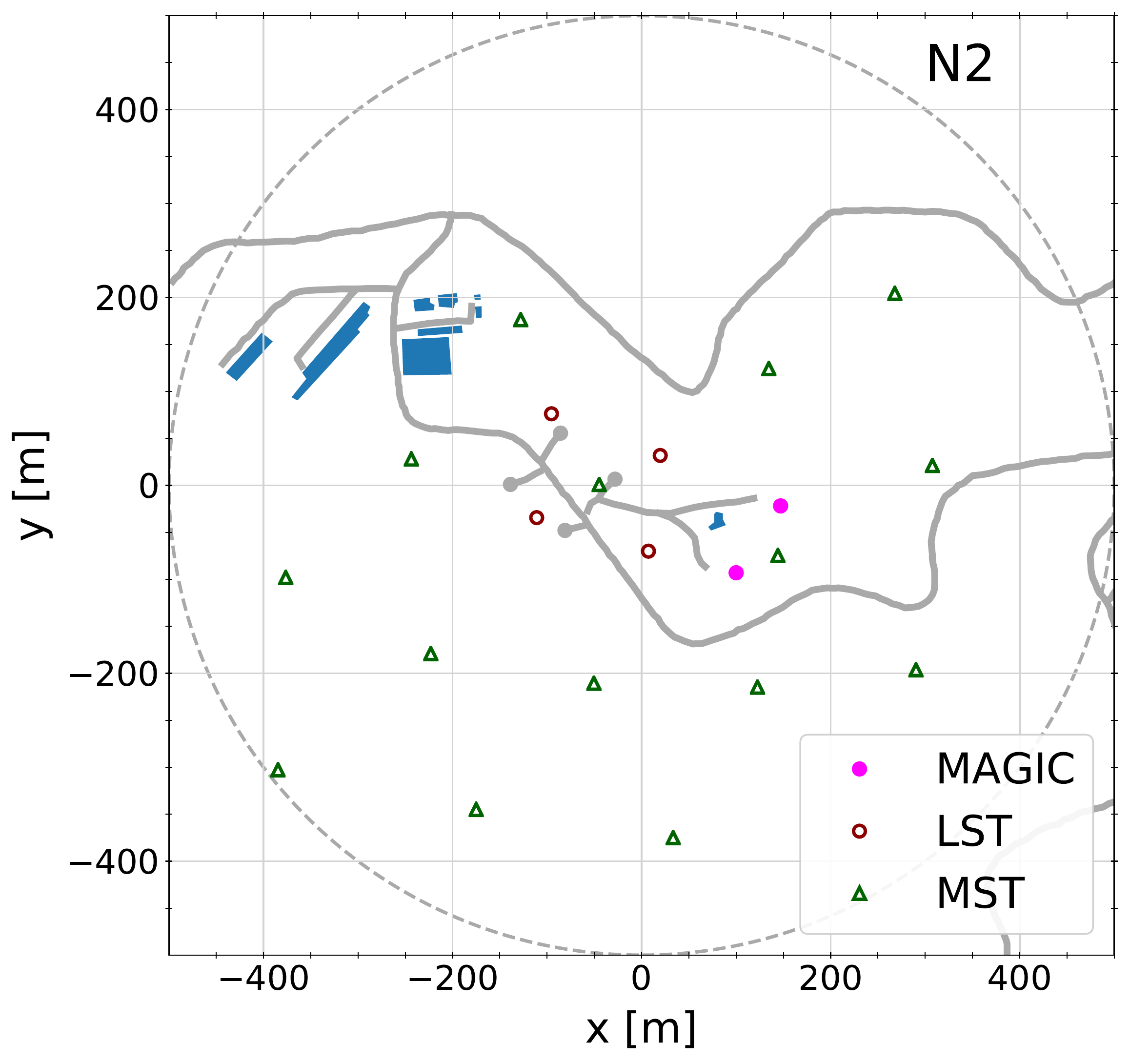}
\includegraphics[width=0.49\textwidth]{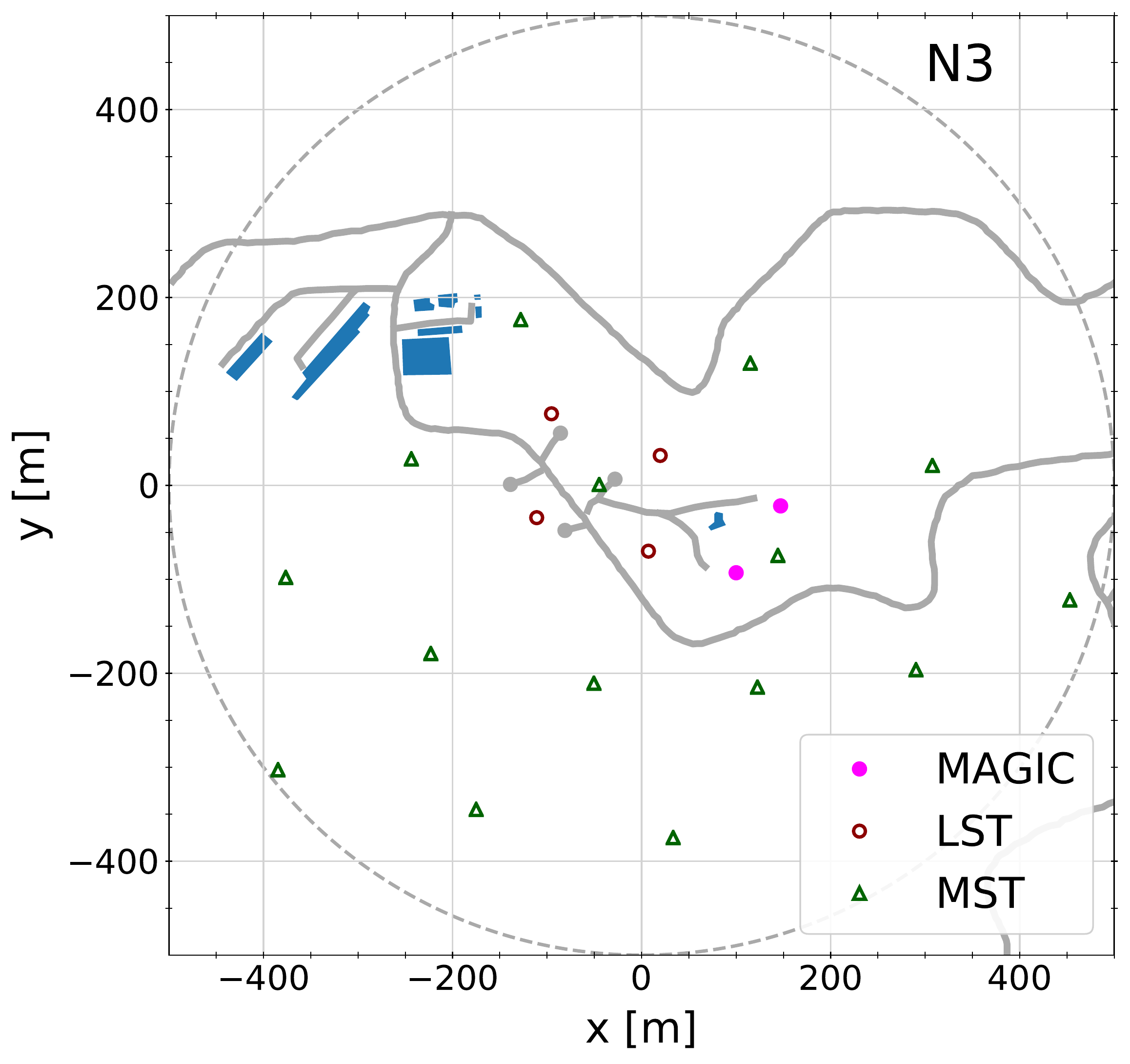}
\includegraphics[width=0.49\textwidth]{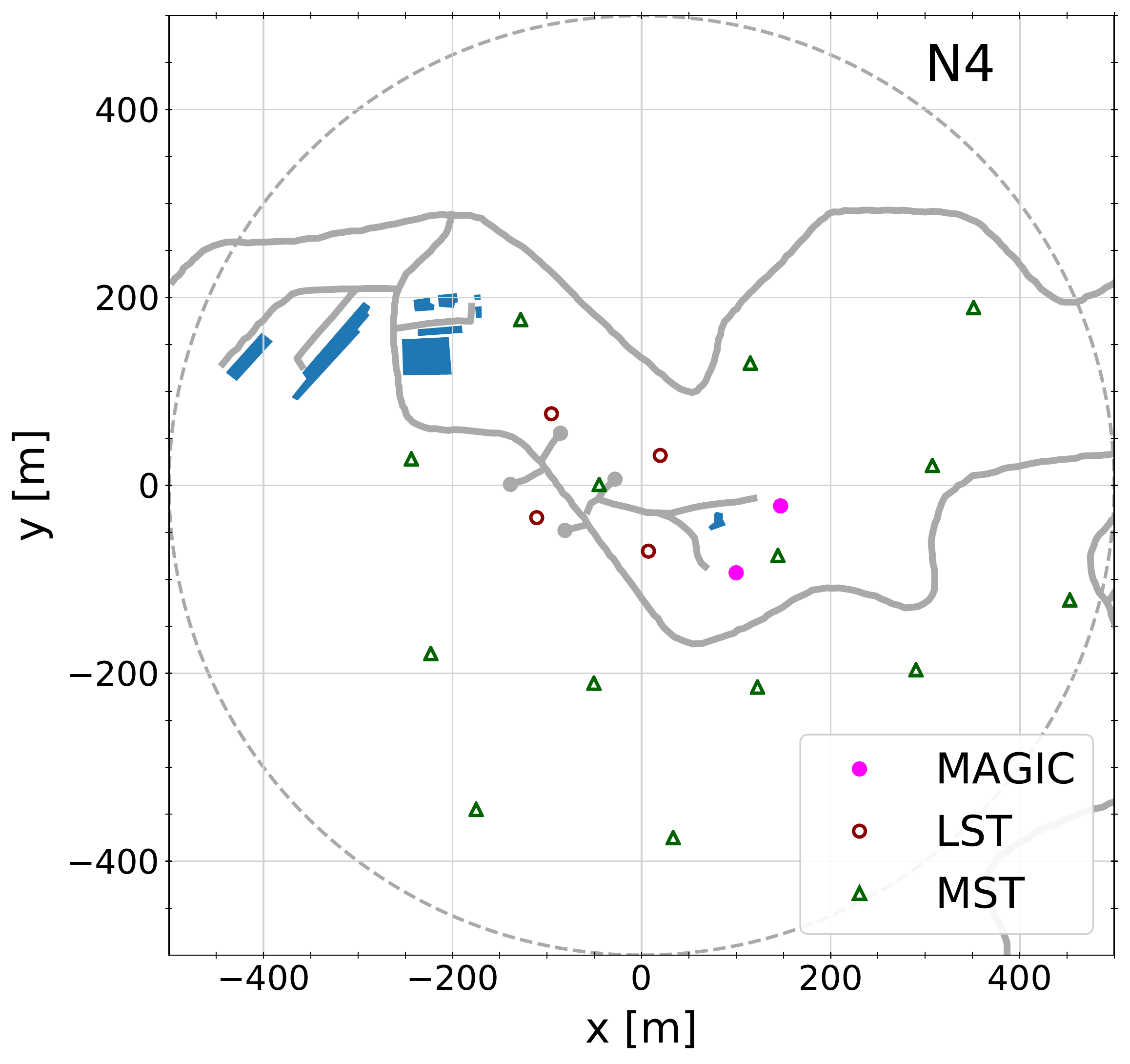}
\includegraphics[width=0.49\textwidth]{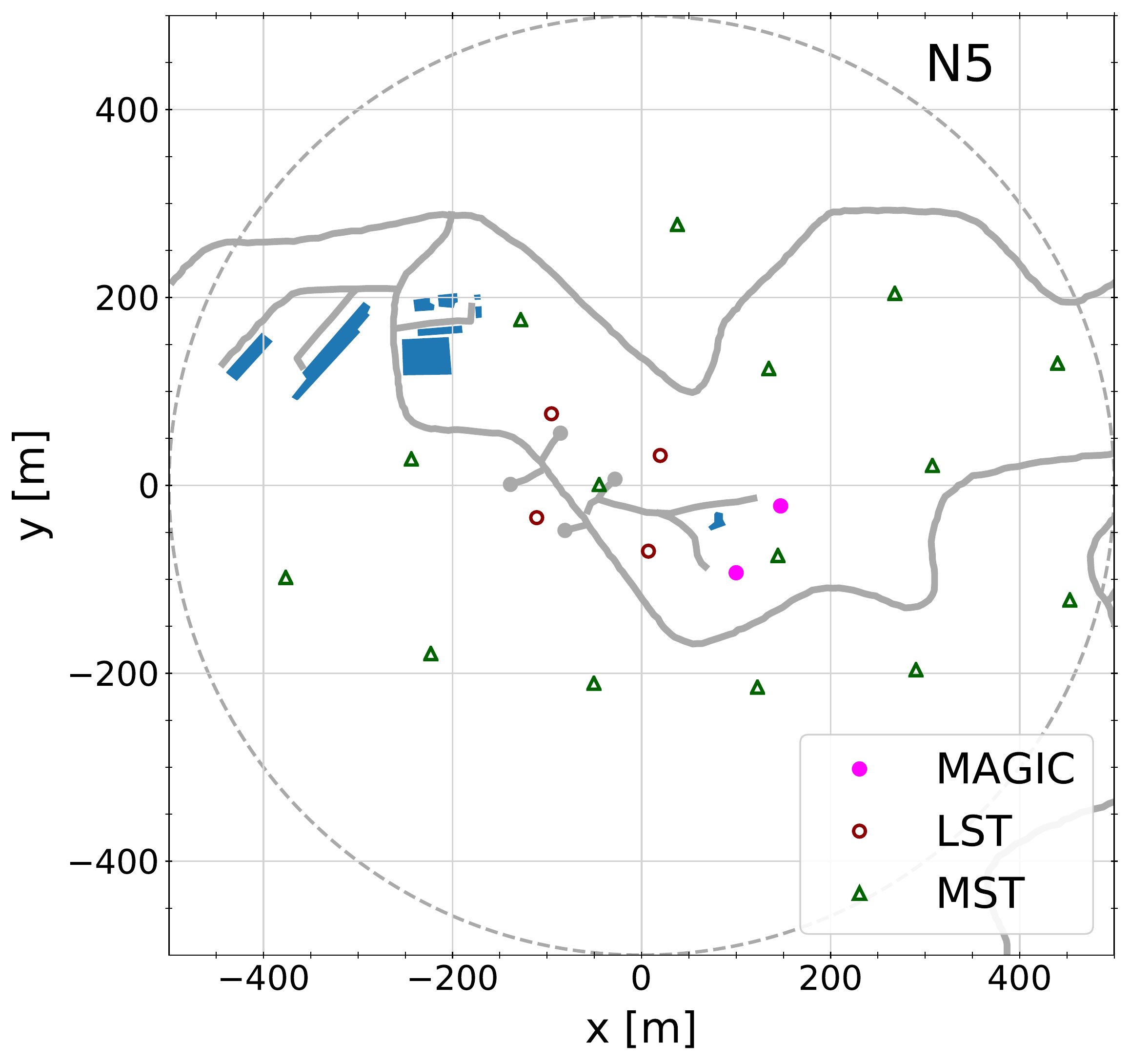}
\includegraphics[width=0.49\textwidth]{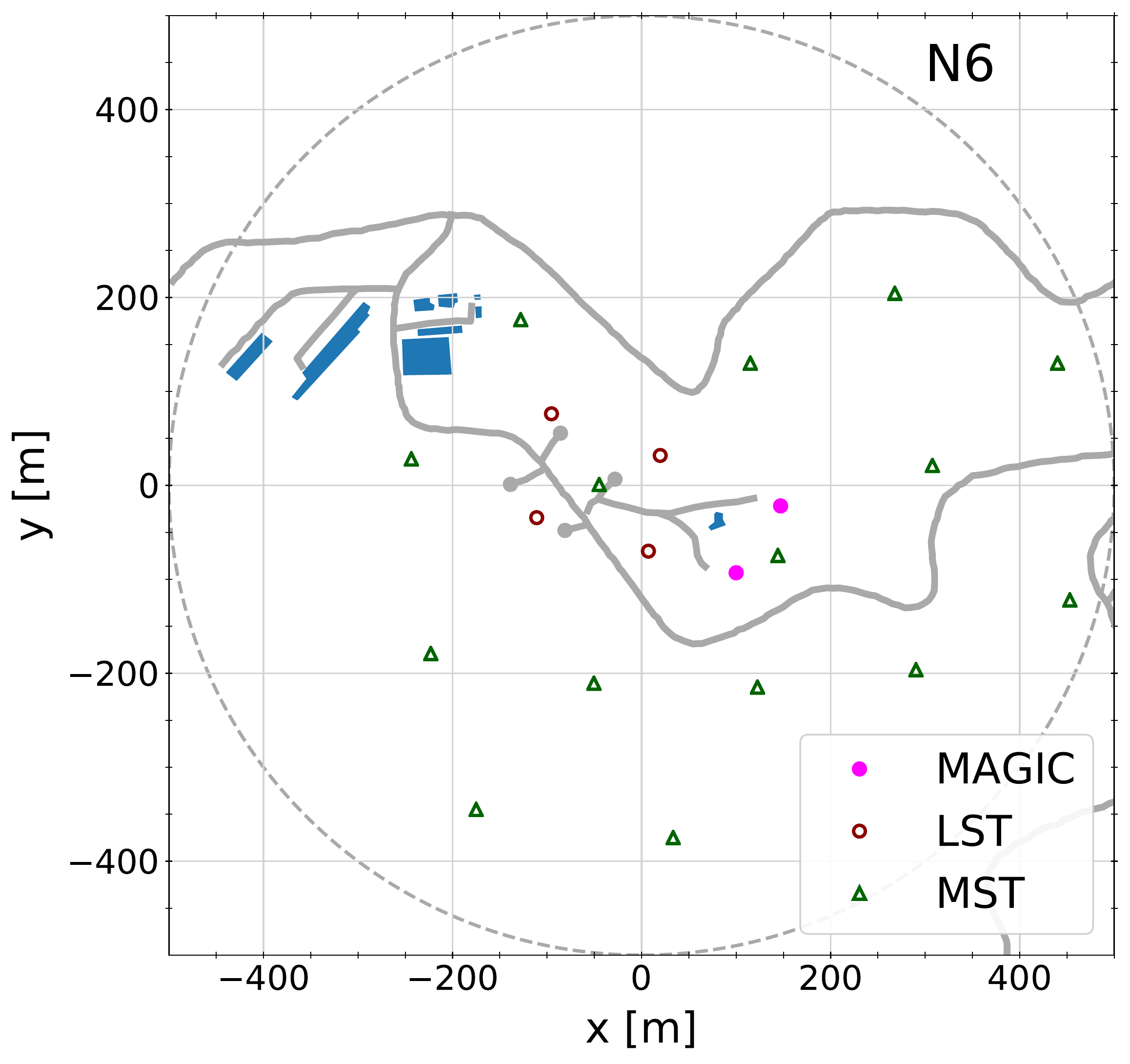}
\caption{Several layouts proposed as baseline arrays for the northern site, together with the position of buildings, roads, and the two MAGIC telescopes. The orography constraints are not shown. The layouts share the LST positions and roughly the same inter-telescope distances between MSTs.}
\label{fig:AL4}
\end{center}
\end{figure}

\begin{figure}[t]
\begin{center}
\centering\includegraphics[width=0.9\linewidth]{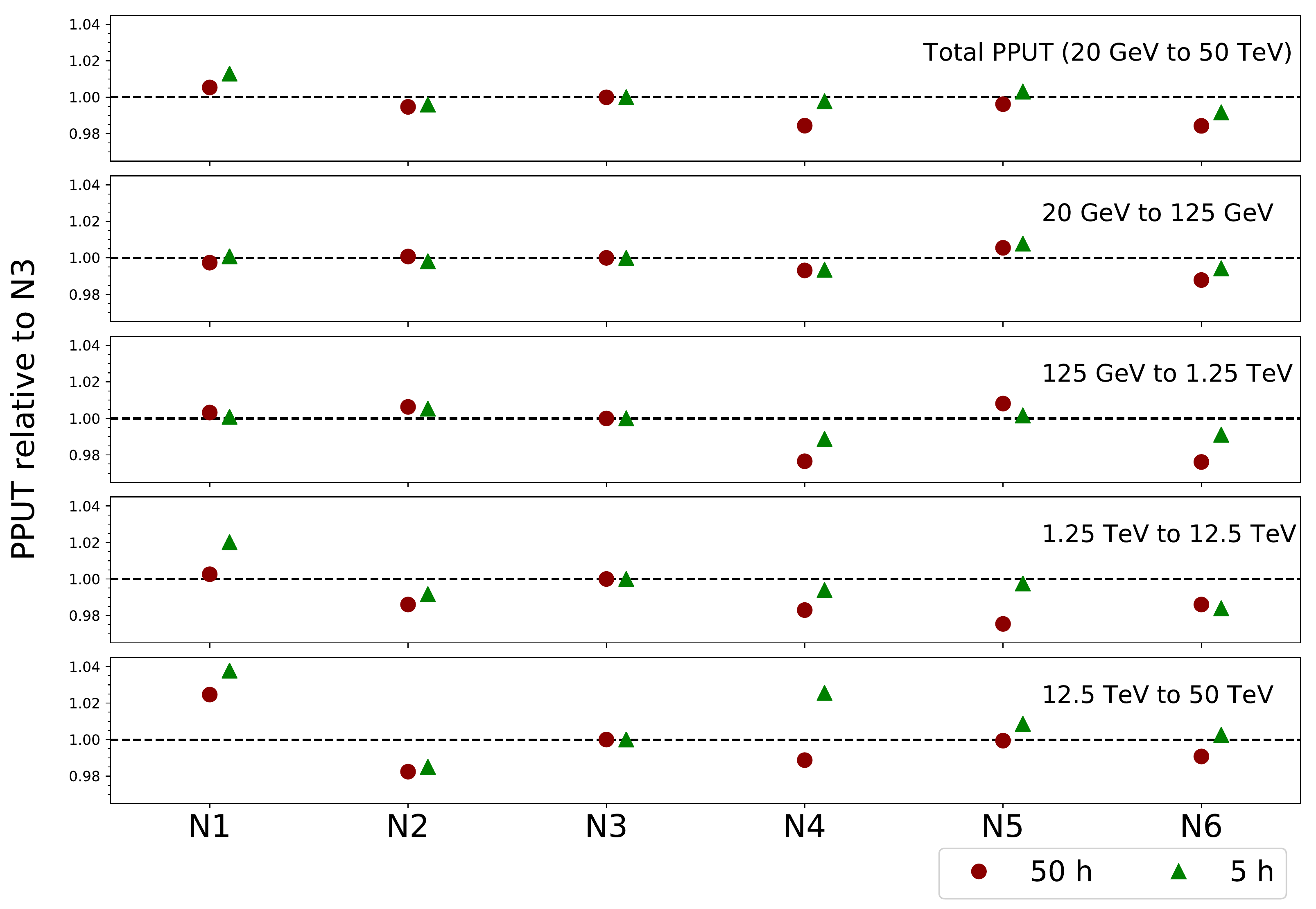}
\caption{Relative PPUT values for several different candidates for the northern layout, relative to ``N3’’. The differences between the layouts are less than 5\%.}
\label{fig:LPPPUT3b}
\end{center}
\end{figure}

The position of the four LSTs was fixed by orography and existing constraints, with LST-1 already under construction. Several solutions are still possible for alternative MST layouts, some of which are shown in Fig. \ref{fig:AL4}, maintaining the same inter-telescope distance. All these alternative layouts achieve similar performance, as shown in Fig. \ref{fig:LPPPUT3b}, while complying with the constraints imposed by the site.

\section{Conclusion}
\begin{figure}[t]
\begin{minipage}{\textwidth}
\begin{center}
\centering\includegraphics[width=0.95\linewidth]{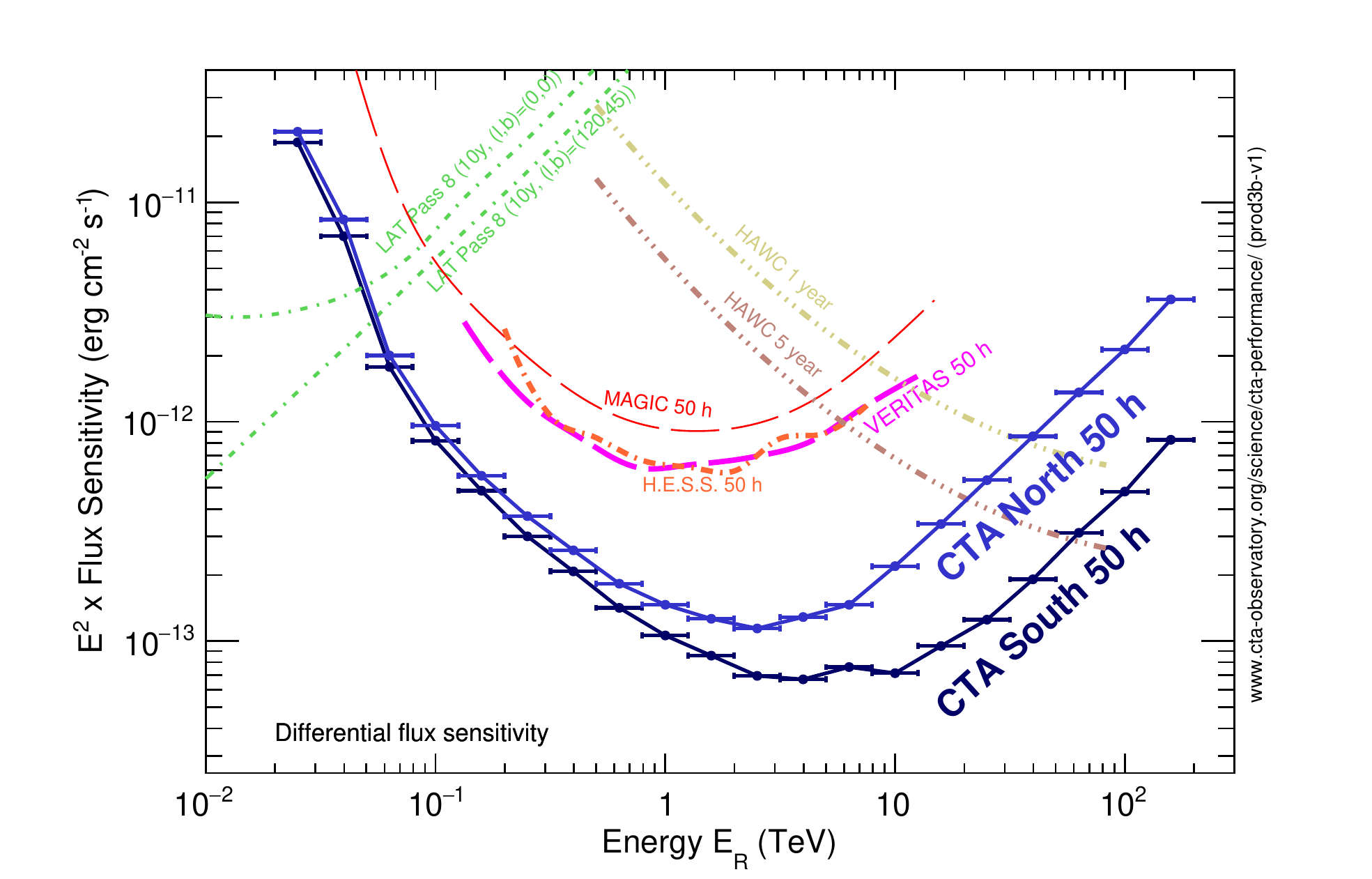}
\caption{CTA differential sensitivity (multiplied by energy squared) 
compared to those of present day instruments (from \cite{publicCTAsensi}): Fermi-LAT \cite{LATSensi}, MAGIC \cite{MAGICSensi1, MAGICSensi2}, H.E.S.S. \cite{HessSensi}, VERITAS \cite{VeritasSensi}, and HAWC \cite{HAWCSensi}}
\label{fig:AllInstr}
\end{center}
\end{minipage}
\end{figure}
The Cherenkov Telescope Array will be the next generation gamma-ray instrument in the VHE range. It will be composed of two separate arrays: the southern observatory will be installed at Paranal (Chile).
The northern array, the construction of which has already started with LST-1, will be built on the island of La Palma (Spain).

These baseline arrays are the result of a concerted effort involving three different large-scale MC productions performed during the last several years. The main purpose of the last large-scale production was to define the final layouts to be constructed in both sites. 
As a result, a single layout (right of Fig. \ref{fig:3HB9Layout}) is proposed for CTA-South. It features a four LST rhombus layout (intermediate step between a square and a double-equilateral triangle), an hexagonal MST layout, and SSTs homogeneously distributed on a circle of about 1.1 km radius. Several similarly performing layouts are instead proposed for CTA-North (Fig. \ref{fig:AL4}).
Given the nearly identical performance of different layouts for CTA-North, the final layout will be fixed based on ease of construction, once a better understanding on the site constraints is attained.

This study shows that the inter-telescope optimum distance of the LSTs is between 100 and 150 m, with a rather flat low-energy performance over these values. The MSTs will provide better performance over the core-energy range of CTA when distributed over a hexagonal grid slightly stretched by applying an azimuthally-symmetric transformation, with inter-telescope distances ranging between 150 and 250~m. The SSTs, present in the southern hemisphere site only, provide better performance in a layout with a strong scaling, with inter-telescope distances ranging between 190 and 300~m.

While the main parameter used in the optimisation is differential sensitivity over the different energy ranges, other considerations were also taken into account. Apart from considering the constraints imposed by the characteristics of the selected sites, minor modifications were applied to the baseline arrays to improve the performance of different staging scenarios (slightly modifying the final LST layout), the cross-calibration between different telescope types, and the stand-alone sub-system performance (mainly by adding SSTs in the inner part of the layout). 

All these layouts comply with the performance requirements imposed by the CTA Consortium for both sites over the full energy range. CTA will outperform present day instruments by more than an order of magnitude in sensitivity in the multi-TeV range, as can be seen in Fig. \ref{fig:AllInstr}. The differential sensitivities presented in Fig. \ref{fig:AllInstr}, together with all the instrument response functions of the proposed baseline arrays, are publicly available \cite{publicCTAsensi} and they were used in the study of \gls{cta} key science projects \cite{ScienceCTA}.

As shown in all the performance comparisons performed throughout this work, the optimisation reaches the few percent level in precision, showing that smaller modifications to these baseline arrays will not lead to significant performance losses. In addition, several different implementations for the SST and MST telescopes were tested and resulted in equivalent conclusions, proving that this optimisation is also valid even if 
different telescope designs undergo minor modifications.

\subsubsection*{Acknowledgments}
We gratefully acknowledge financial support from the following agencies and organizations:

State Committee of Science of Armenia, Armenia;
The Australian Research Council, Astronomy Australia Ltd, The University of Adelaide, Australian National University, Monash University, The University of New South Wales, The University of Sydney, Western Sydney University, Australia;
Federal Ministry of Science, Research and Economy, and Innsbruck University, Austria;
Conselho Nacional de Desenvolvimento Cient\'{\i}fico e Tecnol\'{o}gico (CNPq), Funda\c{c}\~{a}o de Amparo \`{a} Pesquisa do Estado do Rio de Janeiro (FAPERJ), Funda\c{c}\~{a}o de Amparo \`{a} Pesquisa do Estado de S\~{a}o Paulo (FAPESP), Ministry of Science, Technology, Innovations and Communications (MCTIC), Brasil;
Ministry of Education and Science, National RI Roadmap Project DO1-153/28.08.2018, Bulgaria;
The Natural Sciences and Engineering Research Council of Canada and the Canadian Space Agency, Canada;
CONICYT-Chile grants PFB-06, FB0821, ACT 1406, FONDECYT-Chile grants 3160153, 3150314, 1150411, 1161463, 1170171, Pontificia Universidad Cat\'{o}lica de Chile Vice-Rectory of Research internationalization grant under MINEDUC agreement PUC1566, Chile;
Croatian Science Foundation, Rudjer Boskovic Institute, University of Osijek, University of Rijeka, University of Split, Faculty of Electrical Engineering, Mechanical Engineering and Naval Architecture, University of Zagreb, Faculty of Electrical Engineering and Computing, Croatia;
Ministry of Education, Youth and Sports, MEYS  LM2015046, LTT17006 and EU/MEYS CZ.02.1.01/0.0/0.0/16\_013/0001403, CZ.02.1.01/0.0/0.0/17\_049/0008422, Czech Republic;
Ministry of Higher Education and Research, CNRS-INSU and CNRS-IN2P3, CEA-Irfu, ANR, Regional Council Ile de France, Labex ENIGMASS, OSUG2020, P2IO and OCEVU, France;
Max Planck Society, BMBF, DESY, Helmholtz Association, Germany;
Department of Atomic Energy, Department of Science and Technology, India;
Istituto Nazionale di Astrofisica (INAF), Istituto Nazionale di Fisica Nucleare (INFN), MIUR, Istituto Nazionale di Astrofisica (INAF-OABRERA) Grant Fondazione Cariplo/Regione Lombardia ID 2014-1980/RST\_ERC, Italy;
ICRR, University of Tokyo, JSPS, MEXT, Japan;
Netherlands Research School for Astronomy (NOVA), Netherlands Organization for Scientific Research (NWO), Netherlands;
University of Oslo, Norway;
Ministry of Science and Higher Education, DIR/WK/2017/12, the National Centre for Research and Development and the National Science Centre, UMO-2016/22/M/ST9/00583, Poland;
Slovenian Research Agency, Slovenia, grants P1-0031, P1-0385, I0-0033, J1-9146;
South African Department of Science and Technology and National Research Foundation through the South African Gamma-Ray Astronomy Programme, South Africa;
MINECO National R+D+I, Severo Ochoa, Maria de Maeztu, CDTI, PAIDI, UJA, FPA2017-90566-REDC, Spain;
Swedish Research Council, Royal Physiographic Society of Lund, Royal Swedish Academy of Sciences, The Swedish National Infrastructure for Computing (SNIC) at Lunarc (Lund), Sweden;
Swiss National Science Foundation (SNSF), Ernest Boninchi Foundation, Switzerland;
Durham University, Leverhulme Trust, Liverpool University, University of Leicester, University of Oxford, Royal Society, Science and Technology Facilities Council, UK;
U.S. National Science Foundation, U.S. Department of Energy, Argonne National Laboratory, Barnard College, University of California, University of Chicago, Columbia University, Georgia Institute of Technology, Institute for Nuclear and Particle Astrophysics (INPAC-MRPI program), Iowa State University, the Smithsonian Institution, Washington University McDonnell Center for the Space Sciences, The University of Wisconsin and the Wisconsin Alumni Research Foundation, USA.

The research leading to these results has received funding from the European Union's Seventh Framework Programme (FP7/2007-2013) under grant agreements No 262053 and No 317446.
This project is receiving funding from the European Union's Horizon 2020 research and innovation programs under agreement No 676134.

 We would like to thank the computing centres that provided resources for the generation of the Instrument Response Functions \cite{gridAcknowledgement}.

\bibliography{references}

\end{document}